\documentclass[twocolumn]{aastex631}

\begin{document}

\title{Unveiling the Dark Side of UV/Optical Bright Galaxies: Optically Thick Dust Absorption}

\correspondingauthor{Yingjie Cheng}
\email{yingjiecheng@umass.edu}

\author[0000-0001-8551-071X]{Yingjie Cheng}
\affiliation{University of Massachusetts Amherst \\
710 North Pleasant Street, Amherst, MA 01003-9305, USA}

\author[0000-0002-7831-8751]{Mauro Giavalisco}
\affiliation{University of Massachusetts Amherst \\
710 North Pleasant Street, Amherst, MA 01003-9305, USA}

\author[0000-0001-8534-7502]{Bren E. Backhaus}
\affiliation{Department of Physics, 196 Auditorium Road, Unit 3046, University of Connecticut, Storrs, CT 06269}

\author[0000-0003-0883-2226]{Rachana Bhatawdekar}
\affiliation{European Space Agency (ESA), European Space Astronomy Centre (ESAC), Camino Bajo del Castillo s/n, 28692 Villanueva de la Ca\~{n}ada, Madrid, Spain}

\author[0000-0001-7151-009X]{Nikko J. Cleri}
\affiliation{Department of Astronomy and Astrophysics, The Pennsylvania State University, University Park, PA 16802, USA}
\affiliation{Institute for Computational and Data Sciences, The Pennsylvania State University, University Park, PA 16802, USA}
\affiliation{Institute for Gravitation and the Cosmos, The Pennsylvania State University, University Park, PA 16802, USA}
\affiliation{Department of Physics and Astronomy, Texas A\&M University, College Station, TX, 77843-4242 USA}
\affiliation{George P.\ and Cynthia Woods Mitchell Institute for Fundamental Physics and Astronomy, Texas A\&M University, College Station, TX, 77843-4242 USA}

\author[0000-0001-6820-0015]{Luca Costantin}
\affiliation{Centro de Astrobiolog\'ia (CAB), CSIC-INTA, Ctra de Ajalvir km 4, Torrej\'on de Ardoz, 28850, Madrid, Spain}

\author[0000-0002-3331-9590]{Emanuele Daddi}
\affiliation{Universit{\'e} Paris-Saclay, Universit{\'e} Paris Cit{\'e}, CEA, CNRS, AIM, 91191, Gif-sur-Yvette, France}

\author[0000-0001-5414-5131]{Mark Dickinson}
\affiliation{NSF's National Optical-Infrared Astronomy Research Laboratory, 950 N. Cherry Ave., Tucson, AZ 85719, USA}

\author[0000-0001-8519-1130]{Steven L. Finkelstein}
\affiliation{Department of Astronomy, The University of Texas at Austin, Austin, TX, USA}

\author[0000-0002-3301-3321]{Michaela Hirschmann}
\affiliation{Institute of Physics, Laboratory of Galaxy Evolution, Ecole Polytechnique Fédérale de Lausanne (EPFL), Observatoire de Sauverny, 1290 Versoix, Switzerland}

\author[0000-0002-4884-6756]{Benne W. Holwerda}
\affil{Physics \& Astronomy Department, University of Louisville, 40292 KY, Louisville, USA}

\author[0000-0002-6610-2048]{Anton M. Koekemoer}
\affiliation{Space Telescope Science Institute, 3700 San Martin Drive,
Baltimore, MD 21218, USA}

\author[0000-0003-1581-7825]{Ray A. Lucas}
\affiliation{Space Telescope Science Institute, 3700 San Martin Drive, Baltimore, MD 21218, USA}

\author[0000-0001-9879-7780]{Fabio Pacucci}
\affiliation{Center for Astrophysics $\vert$ Harvard \& Smithsonian, Cambridge, MA 02138, USA}
\affiliation{Black Hole Initiative, Harvard University, Cambridge, MA 02138, USA}

\author[0000-0003-4528-5639]{Pablo G. P\'erez-Gonz\'alez}
\affiliation{Centro de Astrobiolog\'{\i}a (CAB), CSIC-INTA, Ctra. de Ajalvir km 4, Torrej\'on de Ardoz, E-28850, Madrid, Spain}

\author[0000-0002-9415-2296]{Giulia Rodighiero}
\affiliation{Dipartimento di Fisica e Astronomia "G.Galilei", Universit\'a di Padova, Via Marzolo 8, I-35131 Padova, Italy}
\affiliation{INAF--Osservatorio Astronomico di Padova, Vicolo dell'Osservatorio 5, I-35122, Padova, Italy}

\author[0000-0001-7755-4755]{Lise-Marie Seill\'e}
\affiliation{Aix Marseille Univ, CNRS, CNES, LAM Marseille, France}

\author[0000-0001-7160-3632]{Katherine E. Whitaker}
\affiliation{University of Massachusetts Amherst \\
710 North Pleasant Street, Amherst, MA 01003-9305, USA}

\author[0000-0003-3466-035X]{{L. Y. Aaron} {Yung}}
\affiliation{Space Telescope Science Institute, 3700 San Martin Drive, Baltimore, MD 21218, USA}

\author[0000-0002-7959-8783]{Pablo Arrabal Haro}
\affiliation{NSF's National Optical-Infrared Astronomy Research Laboratory, 950 N. Cherry Ave., Tucson, AZ 85719, USA}

\author[0000-0002-9921-9218]{Micaela B. Bagley}
\affiliation{Department of Astronomy, The University of Texas at Austin, Austin, TX, USA}

\author[0000-0001-9187-3605]{Jeyhan S. Kartaltepe}
\affiliation{Laboratory for Multiwavelength Astrophysics, School of Physics and Astronomy, Rochester Institute of Technology, 84 Lomb Memorial Drive, Rochester, NY 14623, USA}

\author[0000-0001-7503-8482]{Casey Papovich}
\affiliation{Department of Physics and Astronomy, Texas A\&M University, College Station, TX, 77843-4242 USA} 

\author[0000-0003-3382-5941]{Nor Pirzkal}
\affiliation{ESA/AURA Space Telescope Science Institute}

\begin{abstract}

Over the past decades, a population of galaxies invisible in optical/near-infrared, but bright at longer wavelengths, have been identified through color selections. These so-called optically faint/dark galaxies are considered to be massive quiescent galaxies or highly dust-attenuated galaxies. Having the entire galaxy obscured by dust, however, is likely an extreme case of the much more common occurrence of optically thin and thick absorption coexisting in the same system. With the power of JWST imaging, we are able to spatially resolve massive galaxies at z$\sim$3, accurately model their spectral energy distributions, and identify candidate optically thick substructures. We target galaxies with $\log(M_*/M_{\odot})>$10.3 and 2.5$<z<$3.5, and get 486 galaxies in CEERS and PRIMER fields. Based on excess NIR luminosity, we identify 162 galaxies ($\sim$33\% of the parent sample) as candidate hosts of optically thick substructures. We then carry out spatially resolved SED modeling to explore the physical properties of those dark substructures and estimate the amount of optically thick obscuration. We find that optically thick dust is ubiquitous in normal massive galaxies with a wide variety of SFR and morphology. 10-20\% of the stellar mass/SFR are unaccounted for in our selected galaxies, and the fraction is insensitive to stellar mass or SFR. The dark substructures are generally dustier than the rest of the galaxies and are irregularly distributed, arguing against an obscured AGN as the source of the NIR excess. A correlation between the obscured luminosity and the presence of a recent starburst in the past $\lesssim 100$ Myr is also observed.

\end{abstract}


\section{Introduction} \label{sec:intro}


Since the advent of sensitive near, mid-infrared (NIR, MIR) and (sub)-millimeter imaging, a population of Ultra-Violet(UV)/optically invisible, massive galaxies at high redshifts (z$>$3) has been unveiled. For instance, some red galaxies undetected in deep optical or NIR Hubble Space Telescope (HST) images, have been identified as bright sources at MIR wavelengths and are well detected by the IRAC camera onboard the Spitzer Space Telescope (\citealt{2011ApJ...742L..13H, 2016ApJ...816...84W, 2019ApJ...876..135A, 2021ApJ...922..114S, 2022ApJS..263...38K}). Our current understanding of the cosmic star formation history at z$>$3 relies in measure on Lyman-break galaxies, which are selected from their rest-frame UV/optical emission, and this radiation is attenuated by optically-thin dust, i.e. $\tau_{\lambda}<1$. Therefore, the existence of these so-called HST-dark galaxies (or H-dropouts) indicates that a significant amount of star formation, especially in massive galaxies, is unaccounted for in the cosmic stellar mass and star-formation rate budget at the same redshifts due to the obscuration by optically-thick dust, i.e. $\tau_{\lambda}>1$ at UV/Optical wavelengths (e.g., \citealt{2023ApJ...946L..16P}). 

The origin of these intrinsically red, high-redshift sources is being debated, and it has been proposed that they could be either massive galaxies undergoing very dusty star-formation, or the progenitors of massive quiescent galaxies at early times \citep[e.g.,][]{2018A&A...620A.152F, 2019ApJ...878...73Y, 2019ApJ...884..154W, 2021ApJ...909...23T, 2022ApJ...925...23M, 2024ApJ...962...26G, 2024ApJ...968...34W}.
Recently, the first simulation-based study of this population has been presented by \citet{2024ApJ...961...37C}. Their results support that HST-dark galaxies are high-redshift (z$\gtrsim$4), highly dust-attenuated sources that appear `dark' when viewed along certain lines of sight. In other words, this study suggests that the geometrical distribution of substructures with optically thick absorption is highly inhomogeneous.

Before the launch of the James Webb Space Telescope (JWST), the spectral energy distribution (SEDs) of HST-dark galaxies was poorly constrained at NIR and longer wavelengths by limited spectral coverage and also by the lack of high-resolution information beyond 1.6$\mu m$. Most galaxies are also too faint for ground-based spectroscopy. Now, with the unprecedented capabilities of JWST at infrared wavelengths \citep{2006SSRv..123..485G, 2023PASP..135f8001G}, more optical/NIR-faint, MIR-bright sources have been identified, and the properties of these galaxies have been studied in detail to unveil the nature of their `darkness' \citep[e.g.,][]{2023ApJ...946L..16P, 2023MNRAS.518L..19R, 2023arXiv231107483W, 2024ApJ...961...37C, 2024ApJ...968....4P, 2024ApJ...969L..10P}. A conclusion from these works is that these sources are massive dusty star-forming galaxies at high redshifts, and they are essential contributors to the star formation history (SFH), number density, and assembly history of massive galaxies in the early Universe. The major challenge in studying the properties of these galaxies includes the lack of systematic spectroscopic observations, limited wavelength coverage, and inhomogeneous sample selection. As demonstrated by \citet{2024arXiv240408052B}, spectroscopic data play a significant role in unveiling the nature and physical properties of HST-dark galaxies.

Many existing studies apply integrated color selections to identify optically faint sources, which only catch galaxies whose UV/optical SED is fully obscured by dust
everywhere \citep[e.g.,][]{2020A&A...642A.155Z, 2022ApJ...926..155S, 2023MNRAS.522..449B, 2023A&A...672A..18X, 2023arXiv230902492X}. A situation in which a whole galaxy is optically dark, however, should be relatively rare. It should be more common for galaxies to have regions of optically thin obscuration coexisting with regions of optically thick obscuration (i.e. patchy obscuration and optically-dark substructures), and this should reflect in the spatial variation of their SFHs. In the local Universe, evidence of patchy optically thick clouds has been found in typical disk galaxies \citep[e.g.,][]{2005AJ....129.1396H}. At high redshift (z$>$2), the spatial variation of dust attenuation is also found through spatially resolved measurements of galaxy properties \citep{2023ApJ...946L..16P, 2023ApJ...948..126G}. If this situation is common at high redshift, unless high-resolution and deep far-infrared (FIR) and millimeter data for most galaxies are available, their stellar masses and SFRs (and thus the cosmic densities) could be significantly underestimated, as the hidden optically thick sub-regions completely absorb all the UV and optical photons, which thus cannot be recovered by any extinction correction. In other words, the stellar mass/SFR inferred from FIR data would be higher than that inferred from the UV/optical data. However, it is challenging to quantify and correct for the `missing' stellar mass and star formation because of confusion with the optically thin dust absorption, if only rest-frame UV/optical data is available. The presence of an AGN component in the SED, with its own extinction by dust located both in the nuclear region and in the host galaxy, would further complicate the measure.

The contribution of optically dark galaxies (obscured as a whole) to the cosmic star-formation rate density (CSFRD) is non-negligible, reaching $\Phi \geq (1.9-4.4) \times 10^{-3}~cMpc^{-3}$ or $>$3-6\% of the total CSFRD at z$\sim$3.5 \citep{2024A&A...681L...3M, 2024ApJ...968...34W}. The fraction gets higher for more massive galaxies and at higher redshifts. Thus, these obscured sources can affect the shape of the UV luminosity function (UVLF), especially at the bright end \citep{2024MNRAS.534.2062V}, and also bias our understanding of the mass and size evolution of galaxies. If galaxies with dark substructures are more frequent than `whole' dark galaxies, then the contribution of the partially obscured ones could also be important, perhaps more important. Currently, such contribution is unconstrained, since the frequency and magnitude of spatially segregated optically-thick dust absorption in high-redshift galaxies have never been measured.

In this paper, we quantify the star formation rate and stellar mass unaccounted due to the optically thick dust absorption in massive, $M_*>10^{10.3}$ M$_{\odot}$ galaxies at $z\approx 3$. We also explore the physical properties of the identified optically dark substructures in these galaxies. The rest of the paper is organized as follows. In Section \ref{sec:data}, we describe the observed data and our parent sample selection. In Section \ref{sec:candi}, we develop selection criteria to identify candidate galaxies that potentially host optically thick dust absorption within the parent sample. Then the identification of optically dark substructures and their spatially resolved SED modeling are conducted in Section \ref{sec:dark}. We present our main results in Section \ref{sec:res} and make further discussions in Section \ref{sec:dis}. The conclusions of this paper are summarized in Section \ref{sec:con}.

Throughout this paper, we assume a flat cosmology with $\Omega_M$ = 0.3, $\Omega_\Lambda$ = 0.7 and Hubble parameter $H_0 = 70~km^{-1}Mpc^{-1}$. For SED fittings, we adopt a \citet{2003PASP..115..763C} initial mass function (IMF). All the magnitudes are given in the AB system \citep{1983ApJ...266..713O}.

\section{Data and sample selection}
\label{sec:data}
The data used in this study come from the combination of two JWST surveys, the Cosmic Evolution Early Release Science Survey (CEERS, Program ID 1345, \citet{2023ApJ...946L..13F}, and the Public Release IMaging for Extragalactic Research (PRIMER, GO 1837, PI: James Dunlop).

The CEERS observations target 100 sq. arcmin of the CANDELS \citep{2011ApJS..197...35G, 2011ApJS..197...36K} Extended Groth Strip (EGS) field \citep{2007ApJ...660L...1D}. The dataset includes the JWST/NIRCam (and MIRI for part of the field) imaging as well as the HST/ACS and WFC3 imaging. The detailed description of the NIRCam data reduction and photometry measurements are described in \citet{2023ApJ...946L..12B}. We utilize all data available in the CEERS data release DR0.5 and DR0.6, which includes NIRCam photometry from the F115W, F150W, F200W, F277W, F356W, F410M, and F444W filters, spanning a wavelength range from approximately 1 to 5 microns. The astrometrically aligned HST data include imaging from the ACS F606W, F814W bands, and WFC3 F125W, F140W, and F160W bands (F105W available for some sources), covering 0.6-1.6 $\mu$m. The photometry catalog is generated by Source EXTRACTOR \citep{1996A&AS..117..393B} in two-image mode, using the inverse-variance–weighted sum of the PSF-matched F277W and F356W images as the detection image (see \citealt{2023ApJ...946L..13F}). The MIRI imaging includes eight pointings, four providing deep imaging in F560W and F770W, and four providing contiguous wavelength coverage in F1000W, F1280W, F1500W, and F1800W. The data reduction and photometry measurements are described in \citep{2023ApJ...956L..12Y}.

The PRIMER survey is a deep, wide-area NIRCam+MIRI imaging survey covering 378 sq. arcmin. over the HST COSMOS \citep{2007ApJS..172....1S} and UDS \citep{2007MNRAS.379.1599L} fields. The NIRCam data is reduced from the v7 mosaics using GRIzLI\footnote{\url{https://github.com/gbrammer/grizli}}, and the catalogs are built with the aperture photometry code APERPY\footnote{\url{https://github.com/astrowhit/aperpy}} (see \citealt{2024ApJS..270....7W, 2024ApJ...967L..23C} for detailed settings). We adopt all available data in the internal release V2.0.0. The catalogs include NIRCam photometry from the F090W, F115W, F150W, F200W, F277W, F356W, F410M, and F444W filters, together with HST photometry from the ACS F435W, F606W, F814W bands, WFC3 F125W, F140W, and F160W bands (F105W available for some sources).

The search for partially obscured galaxies requires accurate SED modeling. Generally, the biggest limitations to this procedure come from the inadequacies of the SED modeling templates to capture relevant physical properties, e.g. emission lines and the shape of the continuum, and the uncertainty in the assumption of the dust attenuation curve \citep[e.g.,][]{2013MNRAS.435...87M, 2024ApJ...962...59O}. As described below, we mitigate the impact of these two factors through a specific redshift selection and wavelength coverage.

One of the biggest uncertainties in SED modeling at high redshift when using the reddest bandpasses of NIRCam lies in the variation of polycyclic aromatic hydrocarbon (PAH) features \citep{2007ApJ...656..770S,2023arXiv231007766R}. 
However, at the redshifts of our sample, $z\approx 3$, the strong PAH features are all located outside of the NIRCam band-pass and therefore do not impact our SED modeling. As for the attenuation law, we note that different attenuation curves become highly degenerate at $\lambda\geq 0.67\mu m$ (e.g., \citealt{2003ARA&A..41..241D}). Therefore, we selected the redshift of our sample so that the two reddest bandpasses of NIRCam, our bands for searching optically thick dust absorption, fall in this wavelength range. In this way, different choices of dust attenuation law would not significantly affect our results. These two selection criteria translate into opposing requirements for the redshift range of the sample which, when combined together, result in the optimized redshift selection of $2.5<z<3.5$.

To start defining the sample, we select massive galaxies with stellar mass $\log(M_*/M_{\odot})>10.3$ estimated by \texttt{EAZY}, at redshift $2.5<z<3.5$, and with good photometric coverage (detected in all NIRCam bands with S/N$>$3). This results in a parent sample including 100 galaxies in the CEERS field and 386 in the PRIMER fields (163 in COSMOS and 223 in UDS). Among this sample, 25 galaxies have spectroscopic redshifts measured by JWST/NIRSpec (typically inferred by Lyman-$\alpha$, [OIII], and [CIII] emission lines). We observe high consistency between the photo-z and spec-z for most of these galaxies. The redshift differences are $\Delta z<0.1$ for 18 of the galaxies and $\Delta z<0.5$ for 23 galaxies. We note that these differences are smaller than the width of the redshift selection window.

\begin{figure}
\plotone{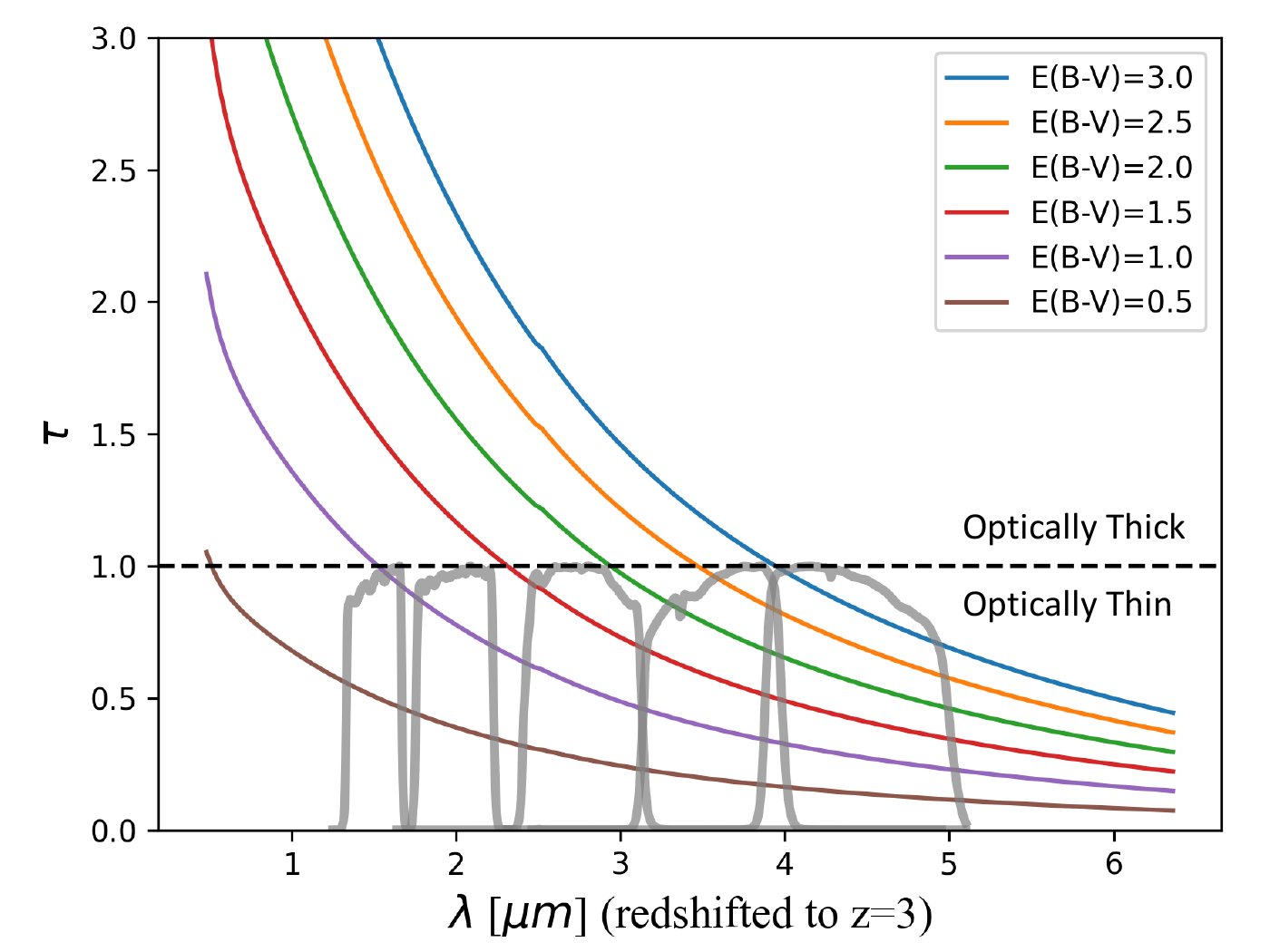}
\caption{The optical depths as a function of wavelengths for different amounts of dust attenuation. The black dashed line ($\tau$=1) shows the optically thin and thick threshold. The throughput of five NIRCam broadband filters (F150W, F200W, F277W, F356W, and F444W) are plotted in gray curves at the bottom. \label{fig:tau}}
\end{figure}

To identify the wavelength range at which, given a value of $E(B-V)$, the transition from (partially) optically thick to optically thin dust absorption takes place, we plot the optical depth $\tau$ as a function of wavelength for a range of $E(B-V)$. Assuming a Calzetti et al. 2000 attenuation curve \citep{2000ApJ...533..682C}, the attenuation is always optically thin ($\tau<1$) in the reddest NIRCam filter, F444W, while in bluer filters, $\tau$ turns bigger than 1 when the dust attenuation gets sufficiently high, indicating that at least part of the absorption becomes optically thick (see Figure \ref{fig:tau}). In this first study, we focus on searching for dust that becomes optically thin in the F444W band, which gives a strict lower limit on the optically thick dust absorption within our selected sample. Note that galaxies with stellar mass and/or redshift outside of the range explored in this work could have different fractions of optically thick obscuration. We leave these cases for future study. Another important assumption that we make throughout the paper is that the intrinsic spectral shapes of the optically thin and optically thick components are the same. While this assumption allows us to estimate the fraction of optically thin vs. thick absorption easily, it might not provide a good description of reality. For example, regions with optically thick dust absorption could originate where there are starbursts, in which case the unobscured intrinsic spectral shape is different than the rest of the galaxy. In this paper, we will neglect this possibility and simply observe that estimates of the obscured stellar mass and star formation rate will be biased low.

\section{Candidates for optically thick absorption}
\label{sec:candi}

\subsection{Selection criterion driven by SED fitting}

For each galaxy in the parent sample, we retrieve all available photometry data from HST and JWST NIRCam and model their SED with \texttt{Prospector} \citep{2021ApJS..254...22J}, excluding the NIRCam F444W filter. This means that potential candidates will be galaxies whose optically-thick dust obscuration becomes thick in the F444W bandpass. \texttt{Prospector} adopts the stellar population synthesis model from FSPS \citep{2009ApJ...699..486C}, and nebular emission model from \citealt{2017ApJ...840...44B}. The parameter settings are described as follows.

We adopt the \citet{2003PASP..115..763C} IMF and the Calzetti's dust attenuation law \citep{2000ApJ...533..682C} in all the fits. The diffuse dust V-band optical depth (the `dust2' parameter in FSPS) is fit with a uniform prior between 0 and 5. The logarithm of stellar metallicity (the `logzsol' parameter in FSPS) is fit with a uniform prior between -4 and 0.4. The redshift and stellar mass are fit with narrow Gaussian priors centered on the measurements from the initial SED modeling made with \texttt{EAZY}. We adopt non-parametric SFH with 7 lookback time bins and the `continuity prior' to control the smoothness in SFR (see also \citealt{2019ApJ...876....3L, 2021ApJS..254...22J, 2022ApJ...935..120J}). The first lookback time bin is fixed at $0<t<30\ \mathrm{Myr}$ to capture recent episodes of star formation, and the last bin is set to be $85\%-100\%$ of the age of the universe at the given redshift. All the intervening bins are evenly spaced in $log(t)$ scale. The sampling is performed with the dynamical nested sampling code $Dynesty$ \citep{2020MNRAS.493.3132S}.

Once we obtain the best-fit SED for each galaxy, we multiply it with the total system throughput in the F444W bandpass\footnote{\url{https://jwst-docs.stsci.edu/jwst-near-infrared-camera/nircam-instrumentation/nircam-filters}} to derive the model-predicted flux in the F444W band ($F_{model}$). If the dust absorption is optically thin in all bands, the $F_{model}$ should match the observed F444W band flux $F_{obs}$ within the uncertainty of the fit. We define the uncertainty as a combination of the model fitting error and the photometric error:
\begin{equation}
    \sigma=\sqrt{\sigma_{fit}^2 + \sigma_{phot}^2 }.
\end{equation}
Then any observed flux excess in the F444W band can be quantified by 
\begin{equation}
    \frac{F_{obs}-F_{model}}{\sigma}.
\end{equation}

A significant flux excess in F444W indicates that the flux in all the bluer bands might have been underestimated, thus identifying the galaxy as a candidate for hosting optically thick dust obscuration. Note that in the selected redshift range no strong emission line is present in the F444W filter. By requiring a significant F444W flux excess with $\frac{F_{obs}-F_{model}}{\sigma}>15$, we pick up 162 candidate galaxies for potential optically thick absorption (36 in the CEERS field and 126 in the PRIMER field), accounting for one-third of the parent sample size. The specific choice of the 15$\sigma$ threshold cut is discussed in Appendix \ref{a:cut}.

\begin{figure*}
\plotone{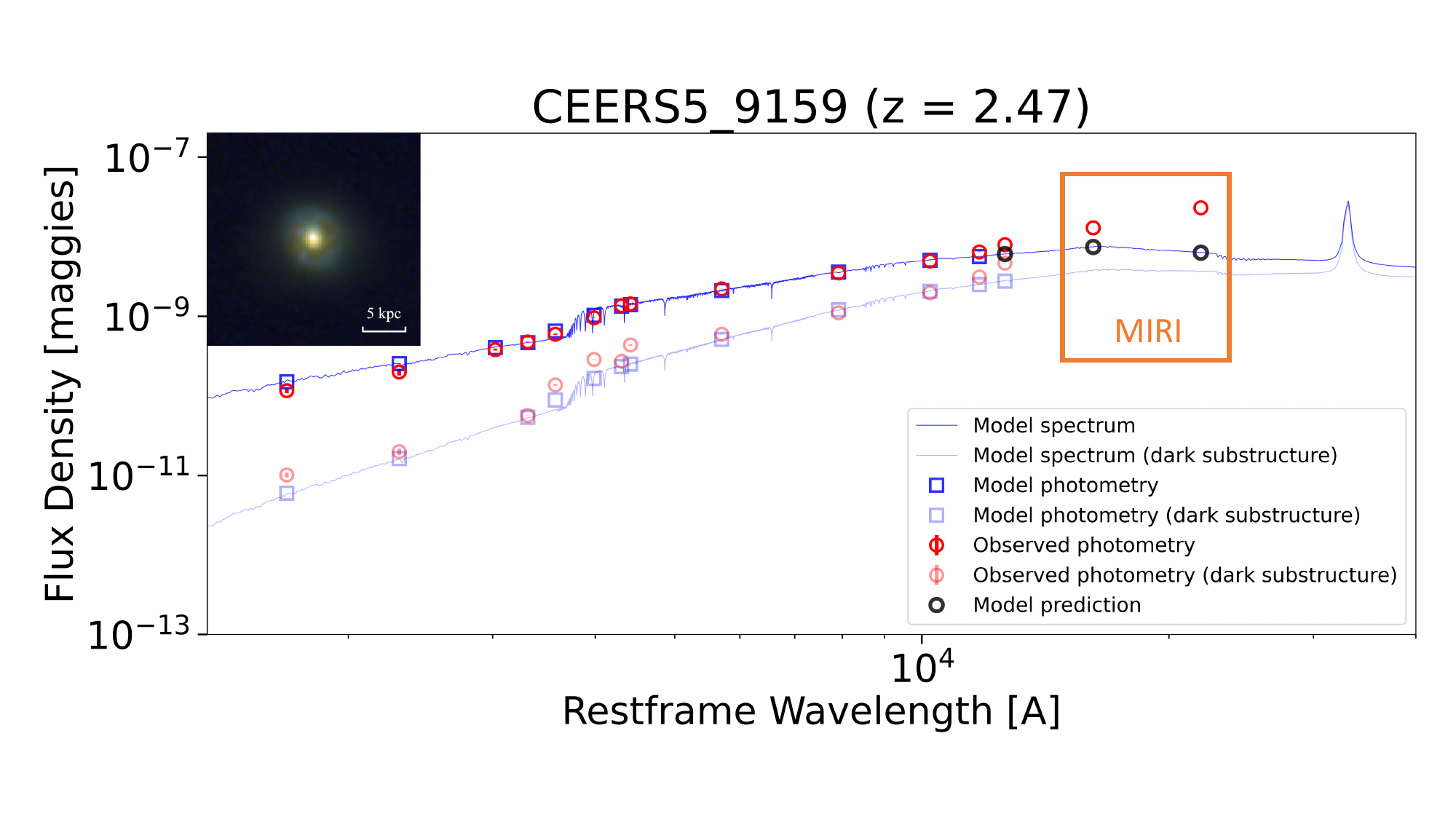}
\caption{The SED modeling for a candidate CEERS galaxy with MIRI data coverage. The best-fit SED for the identified dark substructure is shown with lower transparency. RGB images of the galaxies are shown in the upper left corner, with the F444W for red, the F277W for green, and the F150W for blue. \label{fig:miri}}
\end{figure*}

\subsection{Test of the methodology}
Part of our sample has JWST/MIRI coverage, which gives us the opportunity to explore the validity and robustness of our methodology to identify optically thick dust obscuration inside galaxies. Of our parent sample, 21 galaxies in the CEERS field are covered by MIRI observations, among which 13 are selected as candidate galaxies for optically thick absorption, and are detected in one or more MIRI filters. A small portion of the PRIMER field is also covered by MIRI observations, which will be added to this analysis in the future.

For our candidate galaxies, the excess in F444W is still detected when including available MIRI data in the SED modeling. For 6 of the candidate galaxies, we see significant observed flux excess in all available MIRI bands, where the excess exceeds the F444W band excess in most cases. For 3 of the galaxies, we see clear flux excess in some of the MIRI bands, while the remaining 4 galaxies show no excess in any MIRI band. However, in those 4 cases, the 5.7 and 6.2 $\mu m$ PAH emission lines fall in the wavelength range of the MIRI bands, adding substantial uncertainty to the SED modeling (e.g., \citet{2011A&A...527A.109P, 2024arXiv240303908H}). Therefore, a non-detection of flux excess in MIRI bands does not necessarily mean that the excess in the F444W band is not real, and the possibility of optically thick absorption cannot be ruled out.

As an example, Figure \ref{fig:miri} shows the SED modeling for one of the candidate galaxies. The spectroscopic redshift is available for this source, which substantially reduces the uncertainty in SED modeling. The fitting is based on all bands bluer than F444W, then the fluxes in F444W and the two available MIRI bands (F560W and F770W) are predicted from the best-fit model (black circles). The observed fluxes in all these three bands (red circles) show clear offsets from the model, indicating that a significant amount of optically thick dust absorption turns optically thin at the red part of the SED. The observed flux excesses in the two MIRI filters, along with the excess originally found in F444W, strongly support the existence of optically thick obscuration, and in turn, support our selection criterion. The SED modeling for the candidate optically dark substructure (identified later in Section \ref{sec:dark}) is also shown with lower transparency.

\subsection{Properties of candidate galaxies}
\begin{figure*}
\plotone{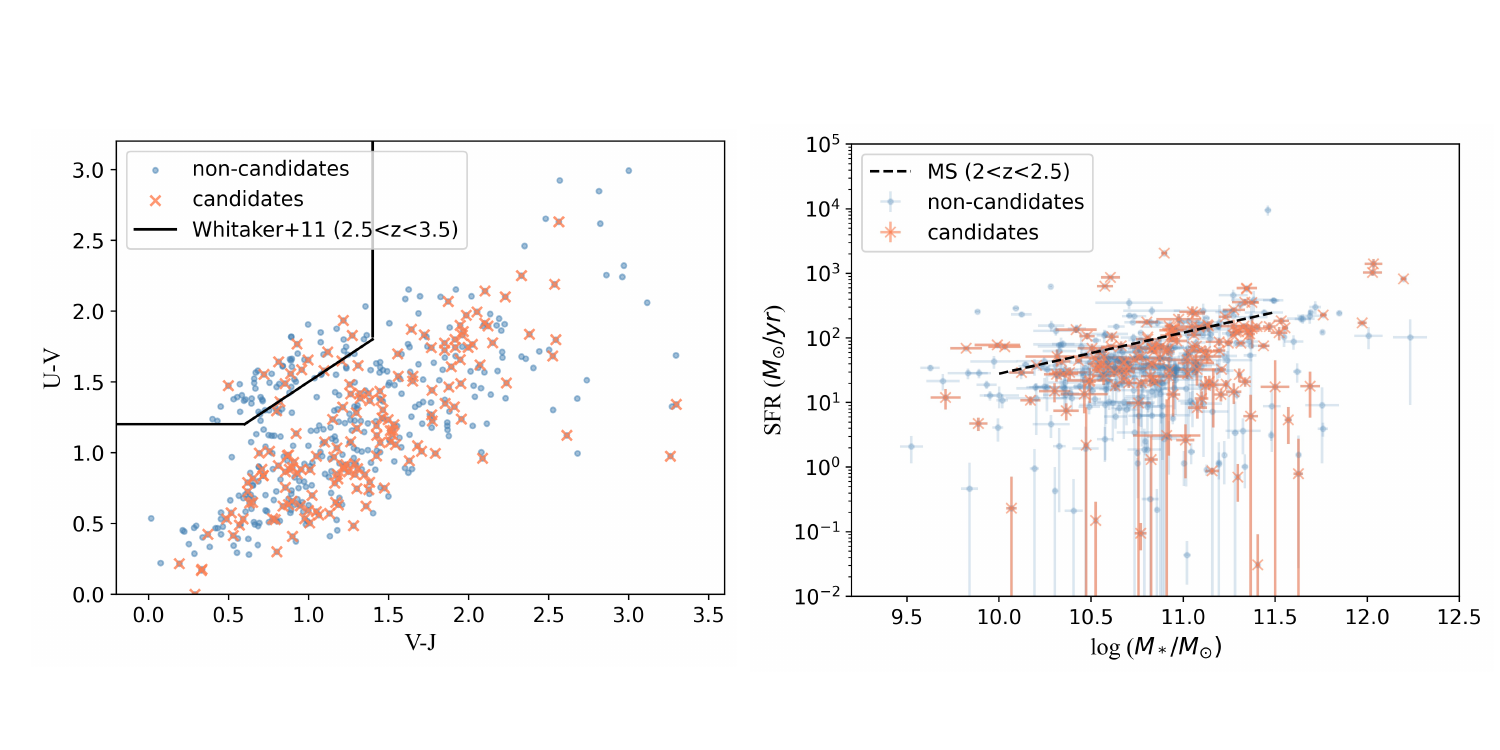}
\caption{The distribution of candidate (orange) and non-candidate (blue) galaxies within our parent sample. The left panel shows the rest-frame UVJ diagram, where the black curve indicates the threshold to separate quiescent and star-forming galaxies given by \citet{2011ApJ...735...86W}. The right panel shows the SFR as a function of stellar mass, with the star formation main sequence \citep{2014ApJ...795..104W} marked in a black dashed line. \label{fig:candi}}
\end{figure*}

Galaxies that are fully dark at optical wavelengths are usually found to be either dust-obscured or quiescent \citep{2024arXiv240408052B}. Since the parent sample is selected purely on the base of stellar mass and redshift, we have a wide spread of galaxies on the star-formation main sequence. Figure \ref{fig:candi} shows the distribution of our candidate and non-candidate galaxies on the rest-frame UVJ diagram (left) and the star-formation main sequence (right). We find that both star-forming and quiescent galaxies can be identified as candidates without any obvious statistical bias. Similarly, we find no significant difference in the distribution of size or morphology (given by \texttt{Galfit} \citep{2002AJ....124..266P} from McGrath et al. in prep.) between candidate and non-candidate galaxies.
This suggests that optically-thick dust absorption is ubiquitous in `normal', massive galaxies at Cosmic Noon. It further implies that the SED with optically thin obscuration of these `partially dark galaxies' is indistinguishable from galaxies that are non-candidates. This also highlights the difficulty of identifying small-scale optically dark patches through the global properties of the integrated galaxy.

\section{Searching for the optically-dark substructures}
\label{sec:dark}

\subsection{Spatial decomposition}
\begin{figure*}
\plotone{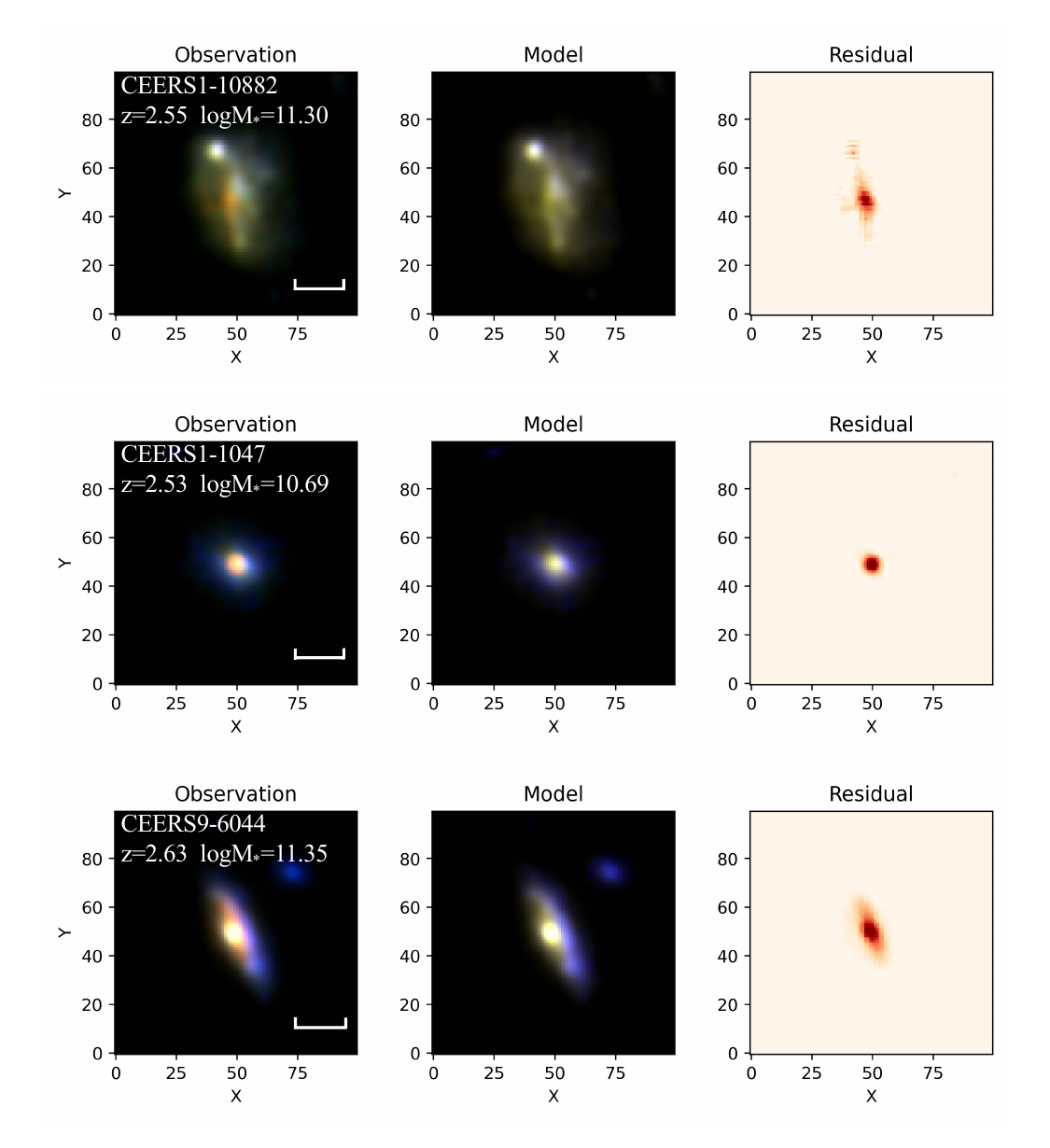}
\caption{Examples of RGB image stamps for three candidate galaxies, with F444W in red, F277W in green, and F150W in blue. The first column shows the observed F444W band image, while the second column shows the model-predicted F444W image. the third column is the residual image obtained by the first column minus the second column. For each galaxy, the catalog ID, redshift, and stellar mass are shown on the label, along with a 5-kpc scale.\label{fig:rgb}}
\end{figure*}

With the sample of the candidate galaxies in hand, we try to locate the optically dark substructures within these galaxies. We adopt different decomposition techniques and, as detailed in Appendix \ref{a:decomp}, we find high consistency among them. To maximize the efficiency in identifying the optically dark regions, we use the F444W band `Obs - Model' residual image to decompose the candidate galaxies and cull the dark sub-regions from the rest. A model-predicted F444W image is derived by normalizing the observed F277W image according to the F444W-to-F277W flux ratio given by the best-fit SED model. Then, we subtract the model-predicted F444W image from the observed F444W image to produce a residual image, which directly reveals the morphology of the observed flux excess in the F444W band. We define the optically dark substructures by applying a 10$\sigma$ threshold cut on the residual image. For visualization, we create RGB images for all the candidate galaxies using the F150W, F277W, and F444W band images, as an analogy to the rest-frame UVJ diagrams. Figure \ref{fig:rgb} shows an example of the RGB image stamps. Visualization for the full catalog will be available online.

\subsection{Digging into the dark substructures}

In the last section, we have identified optically dark substructures for each galaxy in our candidate list. We then measured the fluxes of the identified dark substructures by summing up the pixel values for each band within the defined region. Given the photometry of the dark substructures, we carried out spatially resolved SED modeling with \texttt{Prospector} to explore the physical properties of the dark substructures. Note that this procedure does not prevent that light from optically-thin parts of the galaxies is lumped into the dark regions. Thus, the subsequent comparison will highlight differences between the optically-thin regions and the regions within which the optically-thick dust is embedded.

\begin{figure*}
\plotone{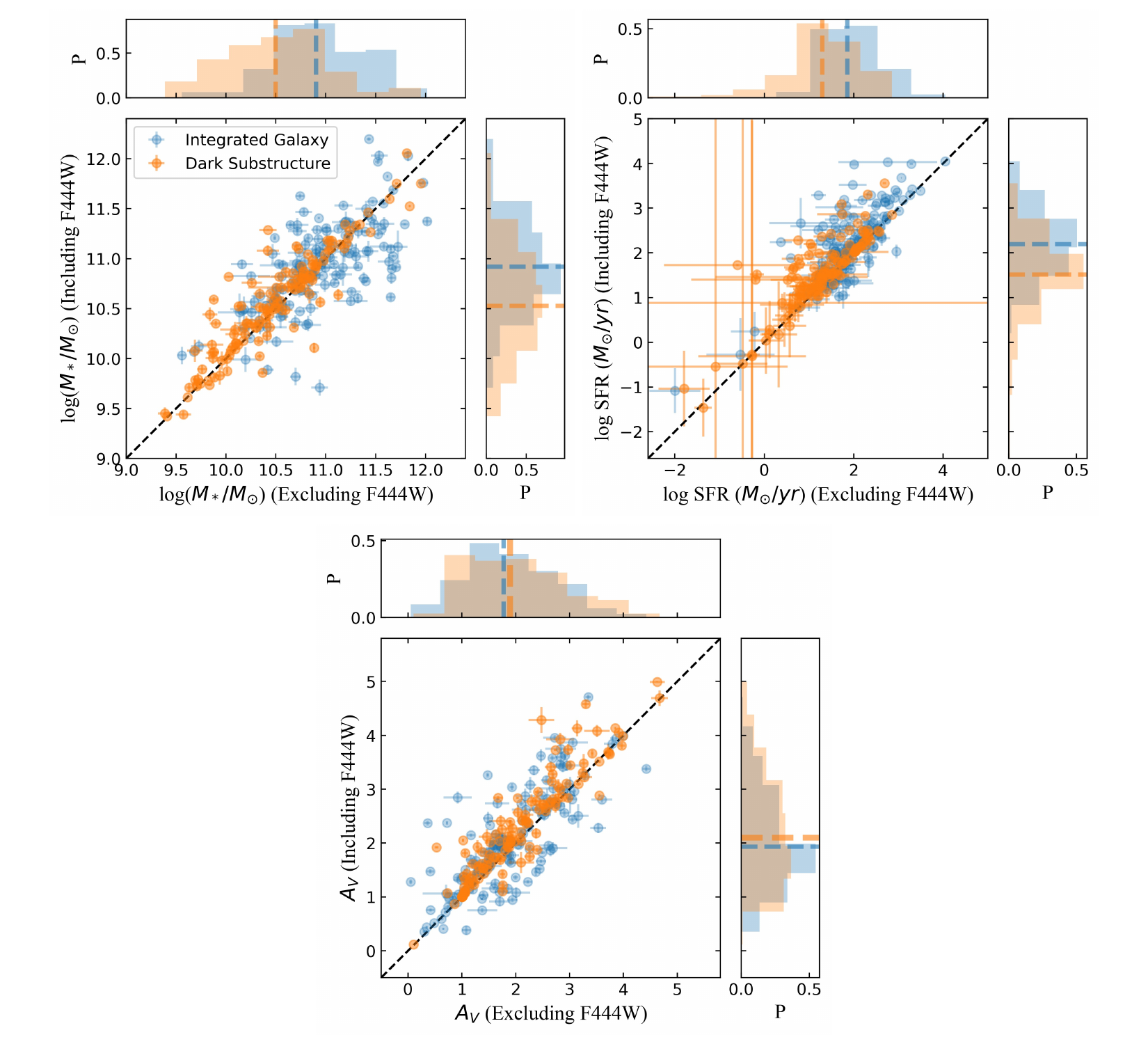}
\caption{The stellar mass, SFR, and dust attenuation of the integrated galaxies (blue) compared with those of the dark substructures therein (orange). The one-to-one relations are plotted in black dashed lines. The number distributions and median values are shown in histograms on both axes.\label{fig:comp_dark}}
\end{figure*}

In this analysis, the redshift is set at the best-fit value given by the previous fit to the integrated galaxy light. Due to the lower S/N ratio in some filters after the photometric decomposition, we reduce the number of lookback time bins from 7 to 5 in the SFH modeling. All the other \texttt{Prospector} parameter settings remain the same. We compare selected galaxy properties of the integrated galaxies versus the dark substructures therein. The comparison of stellar mass, SFR, and dust attenuation are shown in Figure \ref{fig:comp_dark}. 
The stellar mass and SFR within the dark substructures are typically $\sim$one third of the value for integrated galaxies. Moreover, the regions that host dark substructures are generally dustier than the other parts of the galaxies. When comparing the results from the SED modeling with or without the F444W band photometry included, no systematic differences are found for integrated galaxies (scattering symmetrically around the one-to-one relation). For the dark substructures, however, including the F444W band photometry in fitting results in systematically higher stellar mass, SFR, or dust attenuation. Again, this suggests that the optically-thick dust obscuration is not a global feature, instead, it is spatially segregated inside galaxies whose SED is otherwise attenuated by optically-thin absorption. Depending on the magnitude of optically-thick obscuration, it could hardly change the overall properties of the entire galaxy as measured from integrated SED modeling. However, focusing on the identified dark sub-regions is an efficient way to catch the subtle differences in physical properties on or off the dark patches. Similar to optically-dark galaxies, which are known to be dusty massive star-forming galaxies, dark substructures can be understood as dusty, mass-concentrated star-forming regions within common massive galaxies.

\begin{figure*}
\plotone{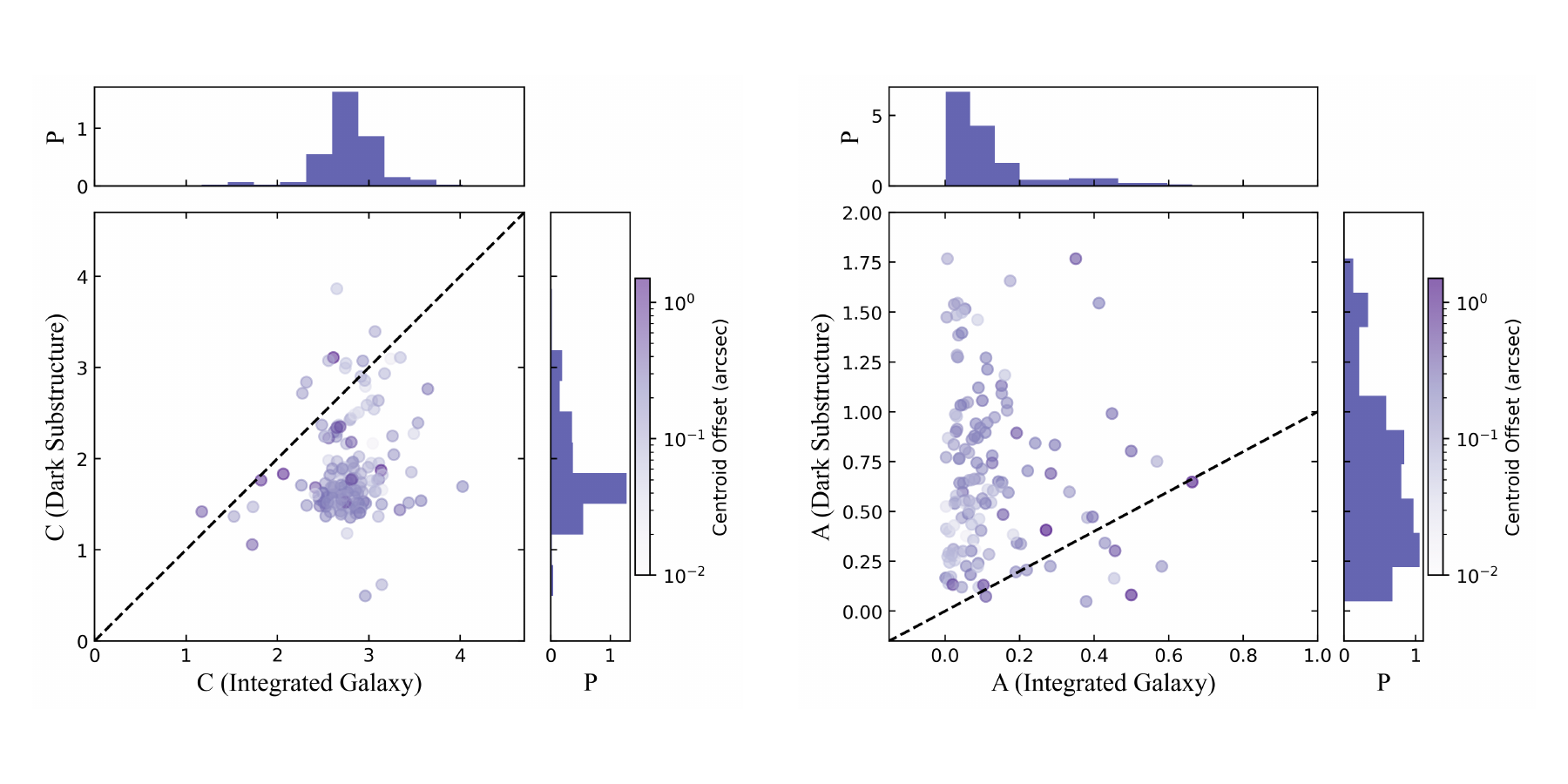}
\caption{The non-parametric morphology measurements of the integrated galaxy and the identified dark substructures. The scatter plot displays color-coded centroid offsets between the two populations, with the black dashed line indicating the one-to-one relation. The number distributions of the two populations are shown in both axes. \label{fig:morph_res}}
\end{figure*}

We see a wide variety of morphologies of the dark substructures within the candidate galaxies. In around half of the cases, the dark sub-region appears to be very compact and located in close proximity to the center of the galaxy (e.g., Row 2 of Figure \ref{fig:rgb}). In other cases, the dark sub-region can be extended and irregular, substantially different from the morphology of the integrated galaxy. The stark contrast between the obscured and unobscured morphologies has also been reported in previous works (e.g. \citealt{2016ApJ...833..103H}).

As a quantitative analysis, we measure the non-parametric morphology of the integrated galaxy as well as the identified dark substructures for all the candidate galaxies using \texttt{statmorph} \citep{2019MNRAS.483.4140R}. The centroid offsets between the entire galaxy and the dark substructures are generally small, with a median value of 0.14$^{\prime\prime}$ ($\sim$1 kpc at z$=$3). While 89\% of the sample show offsets within 0.3$^{\prime\prime}$, we see significant offsets ($>$0.5$^{\prime\prime}$) in 6 galaxies, with the maximum one reaching up to 1.2$^{\prime\prime}$. In Figure \ref{fig:morph_res}, we compare the Concentration and Asymmetry measurements (defined by \citealt{2004AJ....128..163L}) for the entire galaxy and the dark substructures. The dark substructures tend to be more concentrated than the entire galaxy, with a majority showing C$<$2. There is a big spread of asymmetry values for the dark substructures. While the entire galaxy typically has A$<$0.2, the Asymmetry of dark substructures is almost uniformly distributed between 0 and 1, reaching up to $\sim$1.75 in some extreme cases.
Furthermore, galaxies with large centroid offsets tend to be more asymmetric.

\section{Results}
\label{sec:res}

\subsection{Quantifying the obscured star formation}
\label{sec:missing}

\begin{figure*}
\plotone{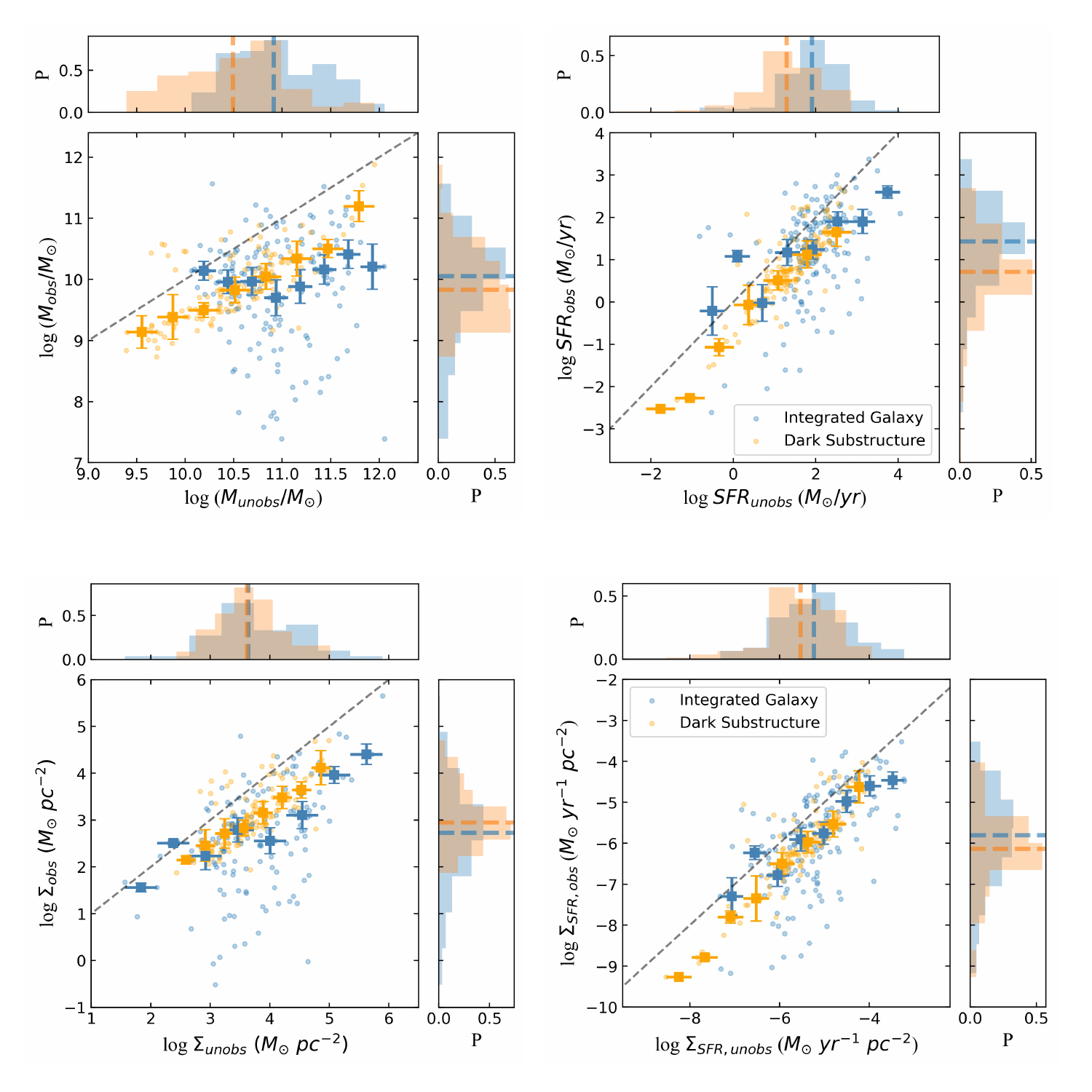}
\caption{Row 1: Comparing the stellar mass (left) and SFR (right) obscured and unobscured by optically thick dust. Blue scattered and binned points represent the integrated candidate galaxies, and orange points represent the identified dark substructures within those galaxies. The dashed line marks the one-to-one relation, and the number distributions and median values are shown in histograms on both axes. The typical relative uncertainties for individual data points are $\delta x=0.7\%$ and $\delta y=1.6\%$ for the left panel, $\delta x=4.3\%$ and $\delta y=25.6\%$ for the right panel.
Row 2: Comparing the stellar mass surface density (left) and SFR surface density (right). The typical relative uncertainties are $\delta x=24.3\%$ and $\delta y=32.4\%$ for the left panel, $\delta x=16.6\%$ and $\delta y=17.4\%$ for the right panel.\label{fig:compare}}
\end{figure*}

\begin{figure*}
\plotone{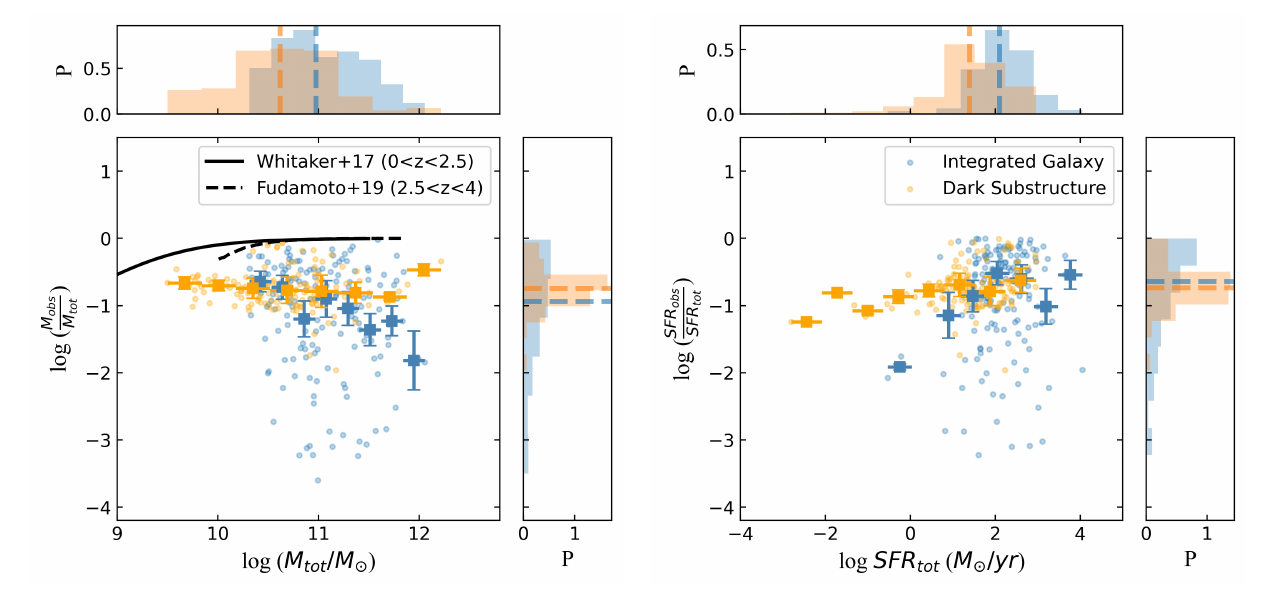}
\caption{The fraction of stellar mass (left) and SFR (right) obscured by optically thick dust. Blue points represent the integrated galaxies, and orange points represent the dark substructures therein. The number distributions and median values are shown in histograms on both axes. The typical relative uncertainties for individual data points are $\delta x=1.3\%$ and $\delta y=51.4\%$ for the left panel, $\delta x=6.5\%$ and $\delta y=36.9\%$ for the right panel. The obscured fraction measurements from two previous works \citep{2017ApJ...850..208W, 2020MNRAS.491.4724F} are shown in black solid and dashed lines in the left panel.
\label{fig:frac}}
\end{figure*}

\begin{figure*}
\plotone{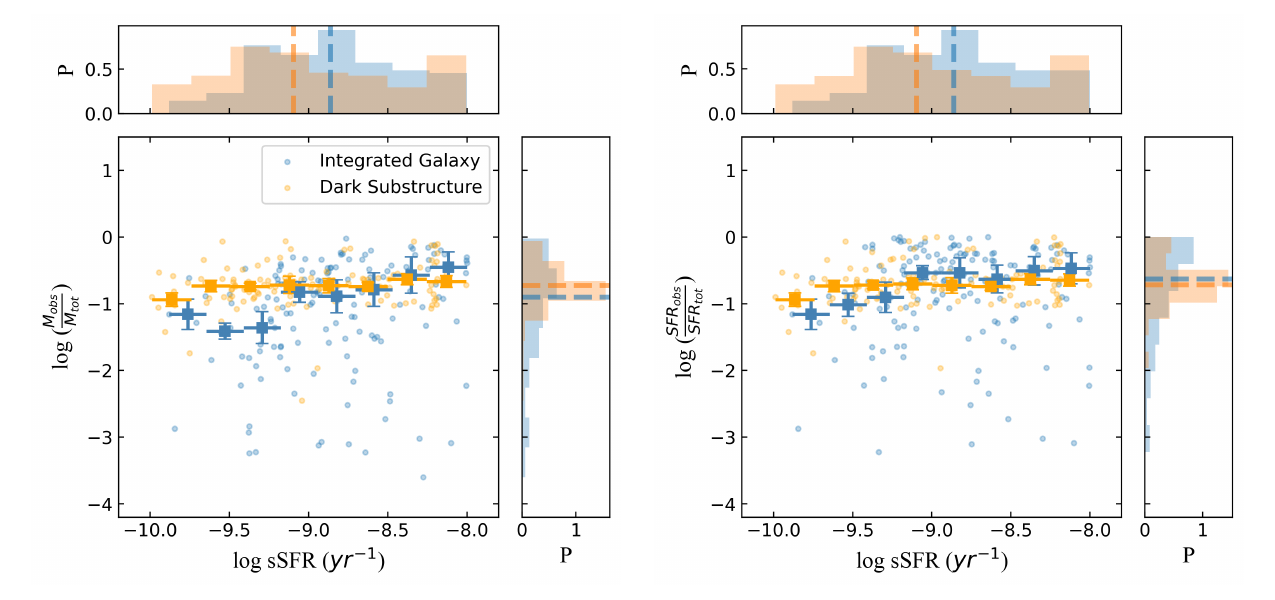}
\caption{The fraction of stellar mass (left) and SFR (right) obscured by optically thick dust as a function of the specific star formation rate (sSFR). The typical relative uncertainties for individual data points are $\delta x=2.2\%$ and $\delta y=51.4\%$ for the left panel, $\delta x=2.2\%$ and $\delta y=36.9\%$ for the right panel. Markers are the same as Figure \ref{fig:frac}.
\label{fig:frac2}}
\end{figure*}

\begin{figure*}
\plotone{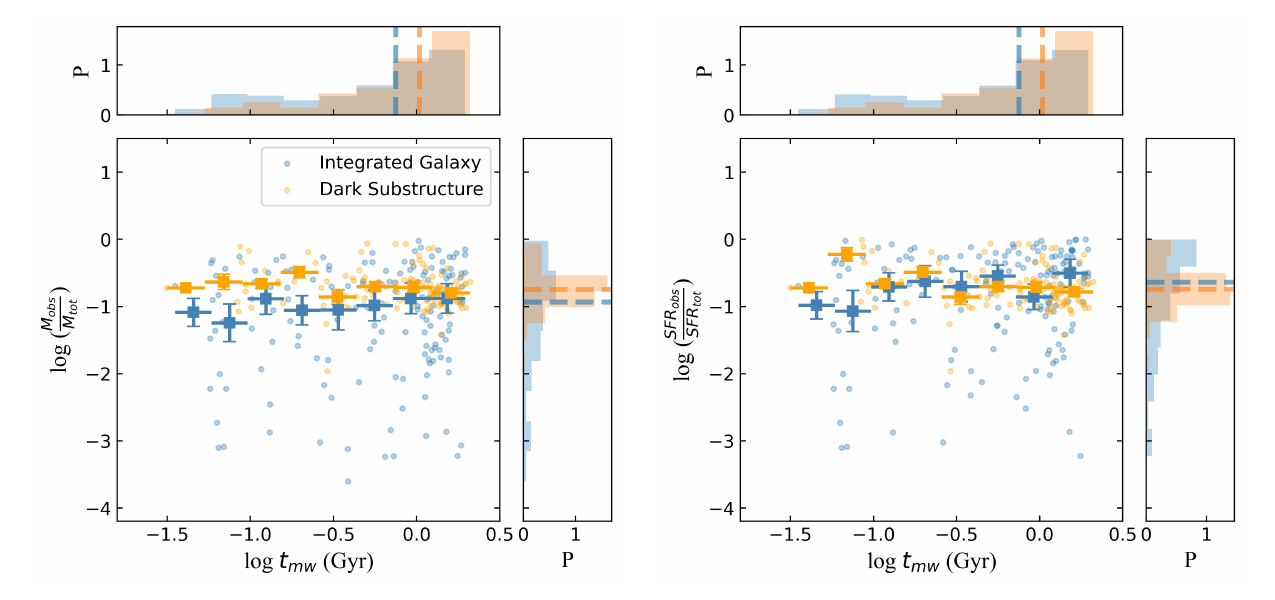}
\caption{The fraction of stellar mass (left) and SFR (right) obscured by optically thick dust as a function of the mass-weighted age ($t_{mw}$). The typical relative uncertainties for individual data points are $\delta x=64.9\%$ and $\delta y=51.4\%$ for the left panel, $\delta x=64.9\%$ and $\delta y=36.9\%$ for the right panel. Markers are the same as Figure \ref{fig:frac}.
\label{fig:frac3}}
\end{figure*}

To simplify the calculation, we assume that the unobscured, intrinsic spectra of the optically thin and optically thick emission are the same. In this way, by subtracting the best-fit model derived from the blue parts of the SED from the total flux estimated by the F444W band photometry, we are able to quantify the contribution of the optically dark component. The dark component can be attributed to the optically thick dust absorption of the stellar light, and also to optically quiet AGNs (e.g., \citet{2024MNRAS.527.12065}). Here we neglect the contribution of any obscured AGN emission, attribute the F444W excess to obscured star formation only, and calculate the `missing' stellar mass and SFR due to the dark component using \texttt{Prospector}. 

The comparison of the stellar mass and SFR unobscured and obscured by optically thick dust is given in Figure \ref{fig:compare}.
The obscured stellar mass is systematically lower than the unobscured stellar mass in our sample. The obscured mass increases with the unobscured mass with a slope close to the one-to-one relation, and apparently flatten at the lower mass end. As for the SFR, the obscured and unobscured values follow the one-to-one relation well, suggesting that they are comparable at all SFRs. Interestingly, this optically-thick fraction is comparable with the obscured fraction of cosmic star formation rate density at z$\sim$3 calculated by $SFR_{IR}/SFR_{UV+IR}$, as suggested by \citet{2023MNRAS.518.6142A}.
A comparison of the obscured and unobscured surface density of stellar mass and SFR is also given in Figure \ref{fig:compare}. While the stellar mass and SFR of the dark substructures are systematically lower than those of the integrated galaxies, their surface densities are comparable. In particular, the median obscured stellar mass density for the dark substructures exceeds the value for integrated galaxies.

In Figure \ref{fig:frac}, we show the optically-thick fraction as a function of stellar mass and SFR respectively. Here the optically-thick fraction is defined as the ratio between the obscured stellar mass (SFR) and the total stellar mass (SFR). The obscured fraction curves from two different studies in the literature are also shown in black lines in the left panel. It is worth mentioning that the definition of the optically-thick fraction presented in this work is different from the obscured fraction defined in those papers, which is calculated as the SFR from IR emission divided by the sum of the SFR from the UV and the IR emission ($\frac{SFR_{IR}}{SFR_{UV+IR}}$). This definition includes obscuration by both the optically thin and optically thick components, leading to a higher fraction than our measurements. Typically, 10-20\% of the stellar mass or SFR is obscured in massive galaxies at z$\sim$3, and the optically-thick fraction is largely insensitive to stellar mass or SFR. Furthermore, we explore the dependence of the optically-thick fraction on the specific star formation rate (sSFR) as well as the mass-weighted age ($t_{mw}$). As shown in Figures \ref{fig:frac2} and \ref{fig:frac3}, such dependence is rather weak, although there is tentative evidence that galaxies with higher sSFRs tend to have more optically-thick obscuration.

\subsection{Dependence on morphology and spectral shape}
\label{sec:morph}

\begin{figure}
\plotone{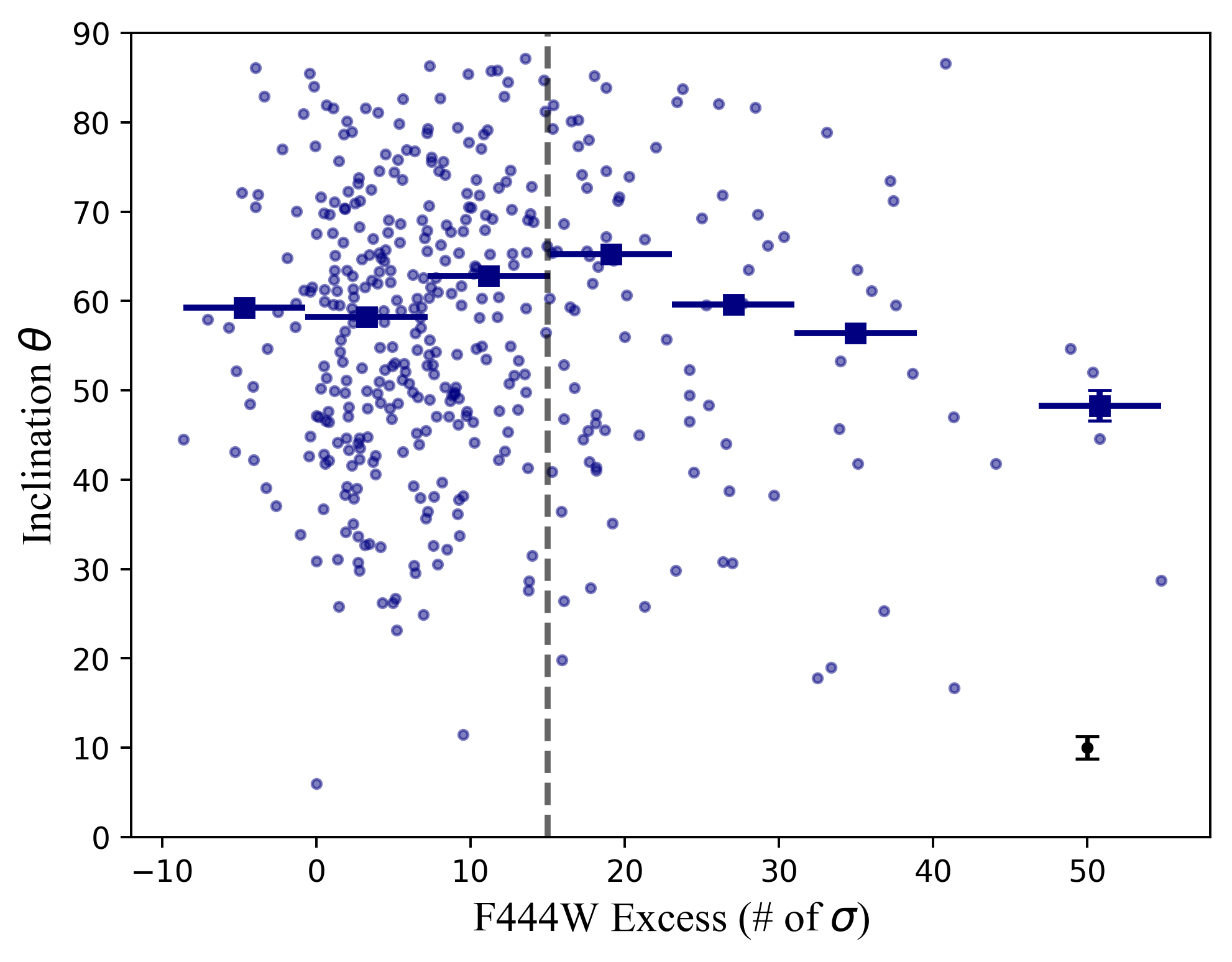}
\caption{The scatter plot and running average of the inclination angles as a function of the F444W band flux excess. The typical error bar of individual data points is shown at the bottom right corner. The vertical dashed line marks our threshold cut for candidate selection. \label{fig:inc}}
\end{figure}

Optically dark galaxies have long been considered to be highly dust-attenuated sources viewed along certain lines of sight, an interpretation which has recently found support from hydrodynamical simulations with radiative transfer modeling \citep{2024ApJ...961...37C}. Thus, inclination would play a key role in producing entirely dark galaxies. It is unclear whether this fact would also be true for the partially dark galaxies probed in this study.

To address this problem, we have calculated the inclination angle of each galaxy in the parent sample based on the axis ratio measurements ($q$) given by \texttt{Galfit} from McGrath et al. (in prep.), using the relation provided by \citealt{1926ApJ....64..321H}:
\begin{equation}
    \cos ^2(i) = \frac{q^2-q_0^2}{1-q_0^2} \quad (q_0=0.20).
\end{equation}
Figure \ref{fig:inc} shows the inclination angles as a function of the observed flux excess in the F444W band. The threshold cut for candidate selection is shown in the gray dashed line. No strong trend is observed for either candidates or non-candidates; candidates might show a very weak trend of increasing excess with decreasing inclination, which would argue that the excess is due to orientation effects since edge-on disks usually appear as more dust-obscured. There is a wide variety of inclination angles among our candidate galaxies (to the right of the dashed line), making them indistinguishable from the non-candidates. This finding lends credence to the locality of optically thick dust absorption within normal massive galaxies. Thus, it seems that the randomly distributed small-scale dark patches cannot dominate the overall properties of the entire galaxy (see also \citet{2022A&A...665A.137S}). The complexity in morphology makes the dependence on inclination angles negligible. 

\begin{figure*}
\plottwo{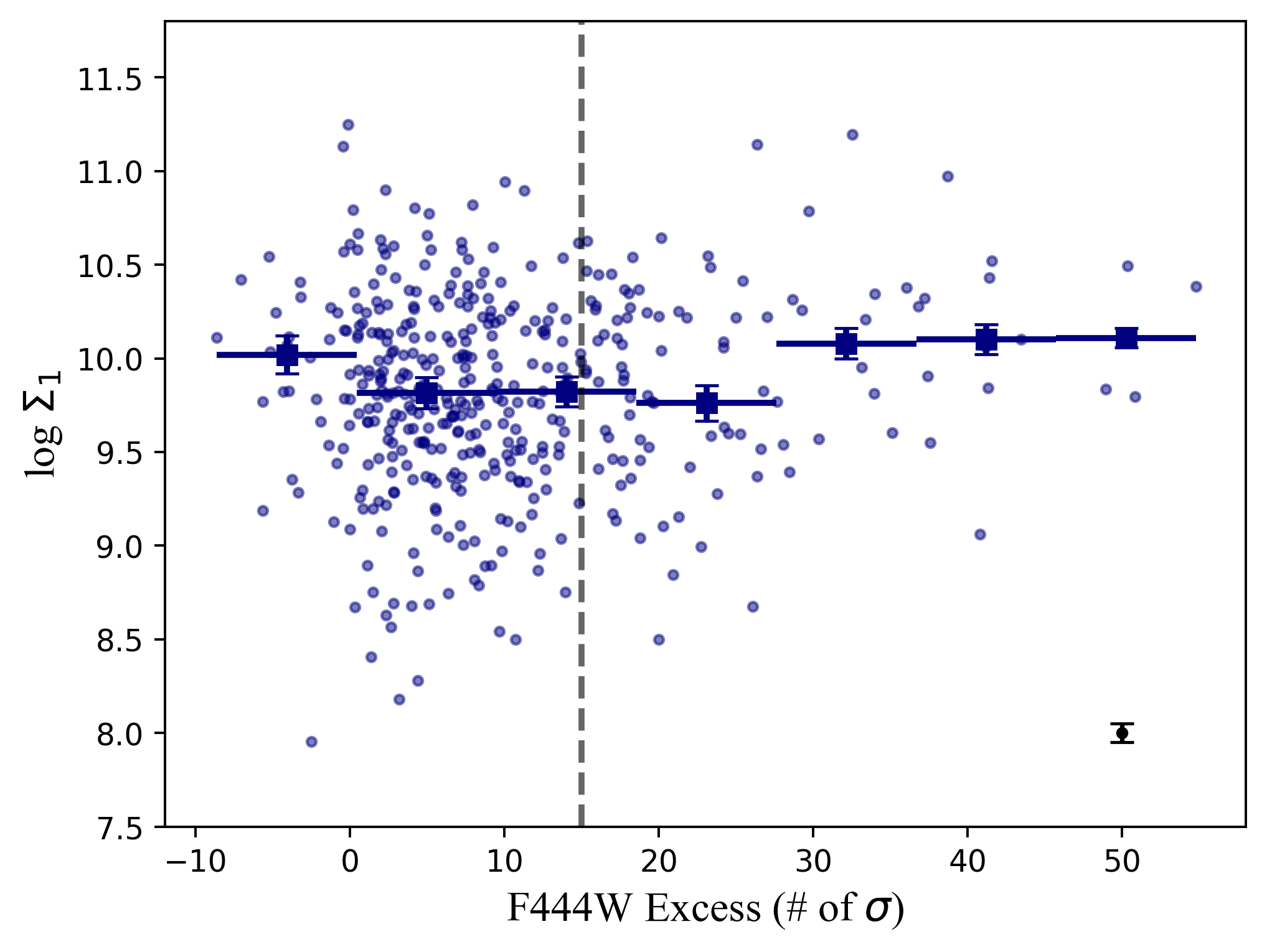}{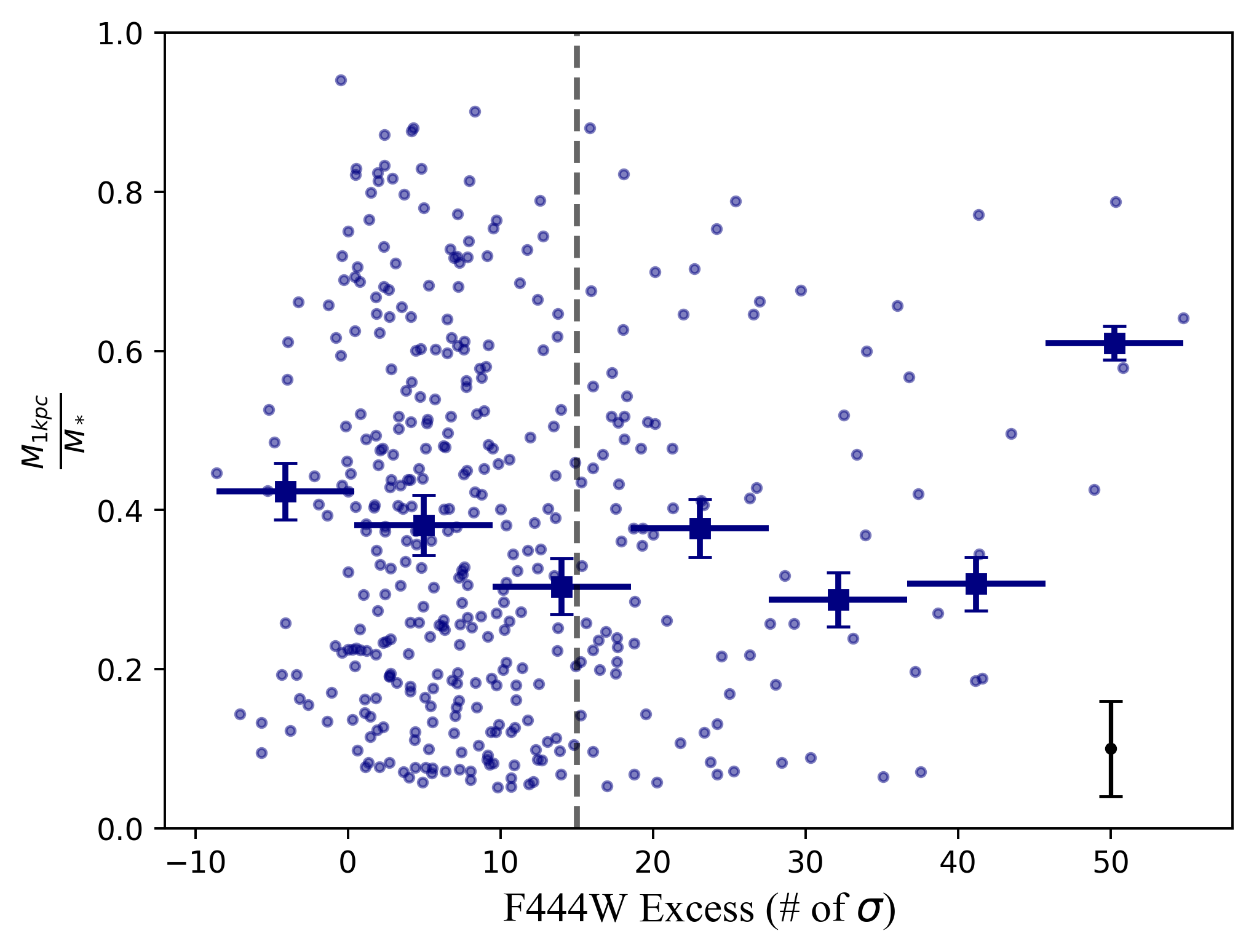}
\caption{The central stellar mass surface density (left) and the normalized central stellar mass (right) as a function of the F444W band flux excess. The typical error bar of individual data points is shown at the bottom right corner for each panel. The vertical line marks the threshold cut for candidate selection.
\label{fig:sigma1}}
\end{figure*}

We also explore the dependence of F444W flux excess on the stellar mass concentration as quantified by  $\Sigma_1$ and $\frac{M_1}{M_*}$, where $\Sigma_1$ refers to the stellar mass surface density within the central 1 kpc, and $\frac{M_1}{M_*}$ is the normalized stellar mass within the central 1 kpc. As shown in Figure \ref{fig:sigma1}, no strong correlation is found between concentration and the presence of optically thick dust absorption. Only in very extreme cases of optically thick absorption (excess$>40\sigma$), the mass concentration tends to be higher.

\begin{figure}
\plotone{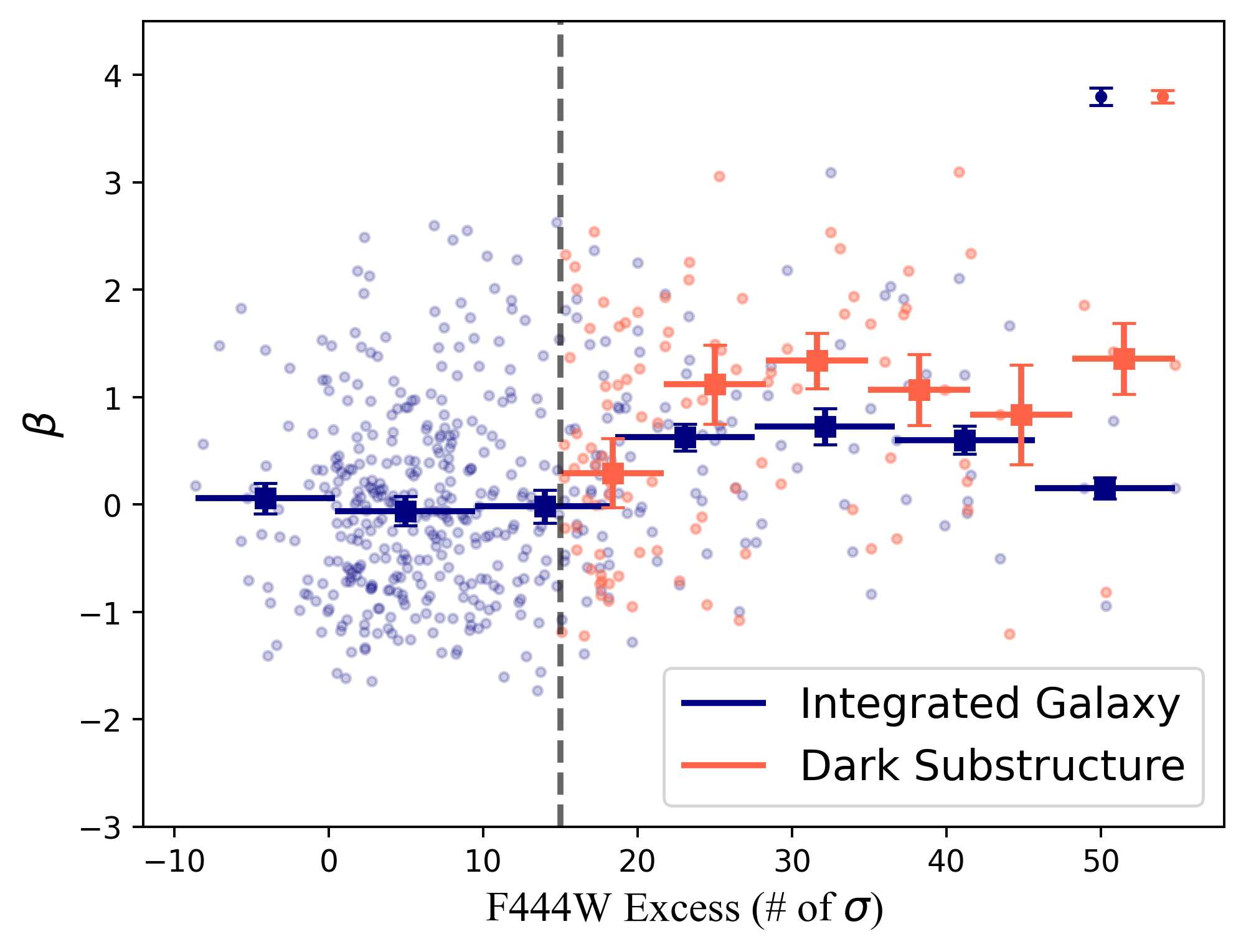}
\caption{The UV spectral slope as a function of the F444W band flux excess. The vertical line marks the threshold cut for candidate selection. For all the candidate galaxies (right to the vertical line), we also plot the UV spectral slopes of the dark substructures in orange, in comparison with the integrated galaxies in blue. The typical error bars of individual data points for both the integrated galaxies and the dark substructures are shown in the upper right corner.
\label{fig:beta}}
\end{figure}

Lastly, we study the dependence of the F444W flux excess on the UV spectral slope $\beta$ for both the integrated galaxies and the dark substructures. We parameterize the shape of the UV continuum as $f_{\lambda} \propto \lambda^{\beta}$ and we fit for $\beta$ in the rest-frame 1268-2580 $\AA$ spectral range of the best-fit SED model. For all of the candidate galaxies, following the image decomposition and the spatially resolved SED modeling, we also derive $\beta$ for the identified dark substructures. Figure \ref{fig:beta} shows the distribution of $\beta$ for both integrated galaxies and the dark substructures therein. Different from the flat distribution centered on $\beta=0$ for the non-candidates, optically thick candidates show a statistically significant increase in $\beta$, reaching $\sim$1 on average. The trend is much more prominent when focusing on the dark substructures instead of the entire galaxy, where the difference is also statistically significant. The Kolmogorov–Smirnov (KS) test between the integrated galaxy data and dark substructure data returns a statistic of 0.833 and a p-value of 0.015, indicating that the two subsets are not drawn from the same distribution. We conclude that galaxies that host dark substructures have, on average, redder UV continuums, with the regions surrounding the substructures being redder than the integrated galaxies as a whole. Finally, we note that while the spectral shape characteristics of the dark regions are diluted when averaged over the entire galaxy, they still remain recognizable in comparison with totally optically thin galaxies.

\subsection{Dependence on the recent SFH}
\label{sec:sfh}

\begin{table*}[h]
\label{table:test}
\centering
\begin{tabular}{lllllll}
& Spearman & Spearman & Spearman & Spearman & KS statistic & KS p-value\\
& coefficient (R1) & p-value (R1) & coefficient (R2) & p-value (R2) &    &   \\
\hline\hline
\multicolumn{7}{>{\centering}m{1\linewidth}}{Integrated Galaxy} \\
\hline
F444W excess $\quad\quad$ & -0.154 & 0.001 & 0.223 & 1.25E-06 & 0.516 & 7.68E-57 \\
$\frac{M_{obs}}{M_{tot}}$ & 0.140 & 0.082 & 0.015 & 0.855 & 0.697 & 2.57E-36 \\
$\frac{SFR_{obs}}{SFR_{tot}}$ & 0.241 & 0.003 & -0.034 & 0.673 & 0.697 & 2.57E-36 \\
\hline
\multicolumn{7}{>{\centering}m{1\linewidth}}{Dark Substructure} \\
\hline
F444W excess & 0.106 & 0.199 & 0.160 & 0.052 & 0.514 & 3.81E-18 \\
$\frac{M_{obs}}{M_{tot}}$ & -0.009 & 0.921 & 0.366 & 1.02E-05 & 0.507 & 1.60E-16 \\
$\frac{SFR_{obs}}{SFR_{tot}}$ & -0.087 & 0.312 & 0.427 & 1.71E-07 & 0.507 & 1.60E-16 \\
\hline
\multicolumn{7}{>{\centering}m{1\linewidth}}{Dark Substructure (small centroid offset)} \\
\hline
F444W excess & 0.173 & 0.146 & 0.244 & 0.039 & 0.431 & 2.26E-06 \\
$\frac{M_{obs}}{M_{tot}}$ & 0.035 & 0.774 & 0.485 & 2.42E-05 & 0.449 & 1.20E-06 \\
$\frac{SFR_{obs}}{SFR_{tot}}$ & 0.030 & 0.808 & 0.567 & 3.77E-07 & 0.449 & 1.20E-06 \\
\hline
\multicolumn{7}{>{\centering}m{1\linewidth}}{Dark Substructure (large centroid offset)} \\
\hline
F444W excess & 0.031 & 0.788 & 0.078 & 0.504 & 0.605 & 2.71E-13 \\
$\frac{M_{obs}}{M_{tot}}$ & -0.043 & 0.724 & 0.280 & 0.020 & 0.580 & 4.59E-11 \\
$\frac{SFR_{obs}}{SFR_{tot}}$ & -0.186 & 0.126 & 0.318 & 0.008 & 0.580 & 4.59E-11 \\
\hline\hline
\end{tabular}
\centering
\caption{Spearman's correlation test and KS test results between the burstiness in recent SFH and the presence of optically thick dust.}
\end{table*}

\begin{table*}[h]
\label{table:ks}
\centering
\begin{tabular}{llllll}
$\quad\quad$  & F444W excess & $\frac{M_{obs}}{M_{tot}}$ &$\frac{SFR_{obs}}{SFR_{tot}}$  & R1 & R2 \\
\hline\hline
KS statistic $\quad$& 0.184 & 0.118 & 0.132 & 0.079 & 0.145  \\
KS p-value $\quad$& 0.152 & 0.664 & 0.529 & 0.973 & 0.406 \\
\hline\hline
\end{tabular}
\centering
\caption{KS test results for the two sub-samples with small and large centroid offsets.}
\end{table*}

Galaxies featuring large amounts of dust obscuration are often found to be starbursting or associated with starburst episodes \citep[e.g.,][]{2020ApJ...895...81R,2021ApJ...923..215C,2022A&A...659A.154I}. Bursts of star formation can be heavily embedded by dust produced by the current and previous generation of stars, though the timescale of the dusty phase remains empirically unconstrained. To gain some insight, we investigate whether there is a connection between recent bursts of star formation and the presence of optically thick dust absorption in the galaxies considered here.

We define two diagnostics to quantify the burstiness, using ratios of SFR in the two most recent bins of look-back time of the SFH from our SED modeling. The most recent burstiness index is defined as the ratio between the average SFR in the recent 0-30 Myr and the average SFR in the recent 30-80 Myr:
$$R1=\frac{SFR_{0-30~Myr}}{SFR_{30-80~Myr}}$$
Similarly, the moderately recent burstiness index is defined as the ratio:
$$R2=\frac{SFR_{0-80~Myr}}{SFR_{80-180~Myr}}$$.
We explore the relationship between $R1$, $R2$, and the presence of optically thick dust as follows. 

\begin{figure*}
\epsscale{0.95}
\plotone{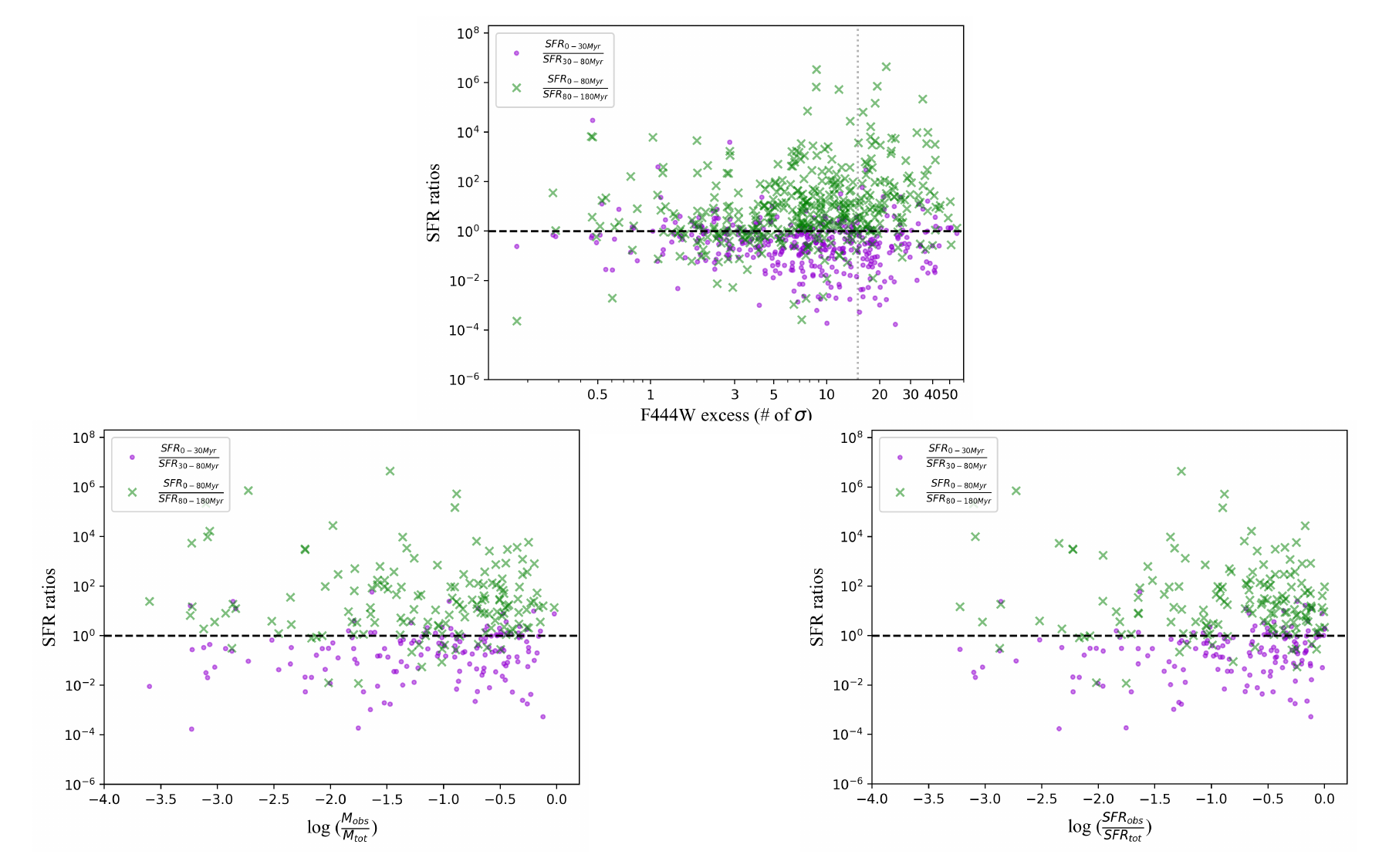}
\caption{The burstiness of the recent SFH as a function of the observed flux excess in F444W (top), and the fraction of stellar mass (left) and SFR (right) obscured by optically thick dust. The two SFR ratios of different timescales are shown in different symbols. The ratio one is marked in a black dashed line.}
\label{fig:burst}
\end{figure*}

The first row of Figure \ref{fig:burst} shows $R1$ (blue points) and $R2$ (orange crosses) as a function of the observed flux excess in the F444W filter. The threshold of F444W excess for candidate selection (15$\sigma$) is marked in a grey dashed line. At the lower F444W excess end, both $R1$ and $R2$ scatter symmetrically around one (black dashed line), but as the F444W excess gets higher, $R2$ are skewed to higher values. Also, a tentative deviation is observed between the distribution of $R1$ and $R2$, with $R2$ typically two times larger than $R1$. The median value of $R1$ for the whole sample is $0.39\pm0.30$, and the median value of $R2$ is $2.73\pm1.32$. The deviation between $R1$ and $R2$ becomes more evident when targeting only the candidate galaxies and using the fraction of stellar mass/SFR obscured by optically thick dust as the y-axis (see the second row of Figure \ref{fig:burst}). For candidate galaxies, the median value of $R1$ is $0.62\pm1.30$, and for $R2$ it is $4.70\pm2.80$, both exceed the values for the whole sample. The majority of $R2$ values are well above one, indicating that galaxies with potential optically thick dust tend to have bursty star-forming activities in the recent $\sim$80 Myr. Evidence of correlation with burstiness is not seen in the most recent $\sim$30 Myr bin, suggesting that the optically thick dust takes some time to be produced and appear in the galaxy SED. We quantify this time interval as $\approx$ 100 Myr.

Since the optically thick absorption is found to be spatially segregated in our candidate galaxies, their integrated SFHs can be contaminated or even dominated by regions with no recent bursts of star formation. To test this possibility, we repeat the analysis based on the SED modeling of identified dark substructures in candidate galaxies. A few examples of the best-fit SFHs of the integrated galaxy as well as the dark substructures are given in Appendix \ref{a:sfh_burst}.
As shown in Figure \ref{fig:burst_dark}, while $R1$ constantly scatters around one, $R2$ typically has values well above one and increases with the strength of the F444W excess, namely the amount of dust obscuration. To quantify the strength of this correlation, we run Spearman's rank correlation tests for all the plots. The results are summarized in Table \ref{table:test}. When using the integrated SFH, we find no strong correlation between $R1$ and the presence of optically thick dust, but there is a positive, relatively weak correlation between $R2$ and the F444W flux excess, with a Spearman's coefficient $\rho\sim$0.2. When fitting the spatially-resolved dark substructures instead of the entire galaxy, a strong positive correlation is observed between $R2$ and the optically-thick fraction for both stellar mass and SFR, reaching $\rho\sim$0.4. The two-sample KS test results are also given in Table \ref{table:test}, suggesting a significant difference between the distributions of $R1$ and $R2$.

Note that the uncertainties in SFH reconstructions from SED modeling propagate to large uncertainties in the measurement of $R1$ and $R2$. Typical SFHs with error bars in this study are shown in Appendix \ref{a:sfh_burst}. The median relative uncertainty for $R1$ is 43\% for integrated galaxies and 39\% for dark substructures. The median relative uncertainty for $R2$ is 67\% for integrated galaxies, and 53\% for dark substructures. The correlations between $R1$ or $R2$ and other galaxy properties can be diluted due to these measurement errors, but the overall trend remains the same.

\subsection{The possible contribution from the AGN.}

\begin{figure*}
\epsscale{1.2}
\plotone{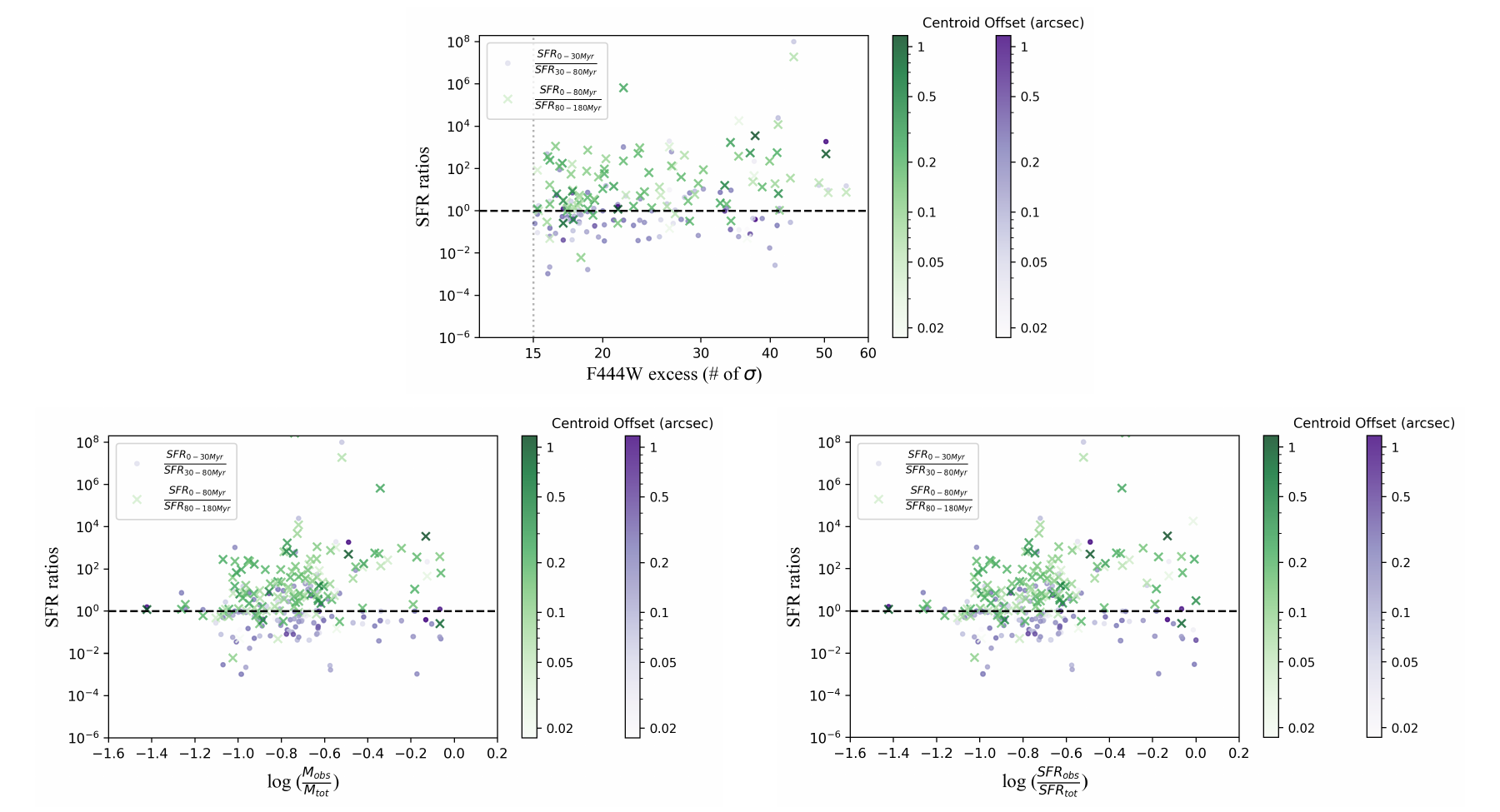}
\caption{The same as Figure \ref{fig:burst}, but derived from the fitting of identified dark substructures. The data points are color-coded by the centroid offset between the entire galaxy and the dark substructure therein.}
\label{fig:burst_dark}
\end{figure*}

As discussed previously, the optically thick absorption can be attributed to not only dusty star formation but also to heavily obscured AGNs. Because AGNs are always located at the center of galaxies, we can use the centroid offset between the entire galaxy and the identified dark substructure to distinguish between these two possibilities. All the data points in Figure \ref{fig:burst_dark} are color-coded by the centroid offsets from null values for both $\Delta X$ and $\Delta Y$ in the galaxies' plane. As the figure clearly shows, there is no significant dependence on the offsets for either axis. As a further step, we divide the candidate galaxies into two sub-samples of equal size based on their centroid offsets. According to the Spearman's correlation tests, the positive correlation between burstiness and the evidence of optically thick dust is slightly stronger for the small centroid offset sample (see Table \ref{table:test}), but the correlation is still observed in both sub-samples. The KS test results between the two sub-samples are given in Table \ref{table:ks}. The p-values for all the given parameters are higher than 0.15, which means we cannot reject the null hypothesis that the two sub-samples are drawn from the same distribution. While both nuclear starbursts and obscured AGNs can contribute to a centrally located optically dark obscuration, the fact that the central and off-central obscuration can be drawn from the same parent distribution argues against the AGN explanation. We conclude that the presence of optically thick dust in our selected sample is more likely related to local bursty star formation activity. Further observations in MIR/FIR wavelengths are necessary to quantify the contribution of nuclear starbursts and obscured AGNs.

\section{Discussion}
\label{sec:dis}

Panchromatic photometry covering from the rest-frame UV to the sub-millimeter has shown that the so-called optically-dark galaxies are massive star-forming galaxies that are highly attenuated by dust obscuration, with large $\tau_{\lambda}$ at UV and optical wavelength. High-resolution imaging with HST, JWST, and ALMA (e.g., \citealt{2023A&A...677A..34G, 2023A&A...677A.172K}) has also provided evidence that the regions with optically thick absorption can be highly inhomogeneous and dependent on the orientation of the galaxy relative to the observer so that galaxies can appear as optically dark when viewed along certain lines of sight and visible when viewed along others (but attenuated by optically-thin obscuration). \citet{sun2023jades} report the JWST observation of a bright sub-millimeter source, which is confirmed to be a heavily dust-obscured star-forming galaxy embedded in an over-dense environment at z$=$5.2. The galaxy has a rest-frame UV-optical counterpart, hence with optically-thin obscuration, which is split into two components due to optically-thick dust obscuration toward the center.

The main result of this paper is that, in addition to the comparatively much rarer optically dark/faint galaxies, optically dark substructures also exist and appear to be relatively common among normal, star-forming galaxies at cosmic noon. Several recent studies have found spatial offsets, of the order of a few kiloparsecs, between the stellar emission, the ISM gas emission, and the dust emission in normal galaxies at high redshifts \citep[e.g.,][]{2018MNRAS.478.1170C, 2019ApJ...881..124M, 2022MNRAS.510.5088B}. According to \citet{2024arXiv240207982K}, around 30\% of the Main-Sequence galaxies at redshift $z=$4-6 display offsets greater than the median by more than 3$\sigma$ significance. These offsets can have a number of physical mechanisms behind their existence, including complex dust geometry, strong stellar and AGN feedback, large-scale gas flows, or the mixture of ISM due to turbulence.

Similar to optically dark galaxies being recognized as massive dusty star-forming galaxies, dark substructures can be understood as dusty mass-concentrated star-forming regions within common massive galaxies. Through ALMA observations, \citet{2023ApJ...948L...8R} have identified dust-attenuated star-forming substructures $\sim$2 kpc away from the galactic center in a typical star-forming galaxy at z$\sim$2.7 (see also \citet{2018A&A...616A.110E}). Similar substructures should exist in many other high-z galaxies. However, it is expensive and unrealistic to expand the search for dark substructures through high-resolution FIR and Submillimeter observations. In this study, without any FIR data, we have used NIRCam photometry to conduct a search in massive galaxies at z$\sim$3 looking for emission that becomes optically thin at around 1$\mu$m and optically thick at shorter wavelengths. In this way, we have identified a sample of partially dark galaxy candidates through state-of-the-art SED modeling. Considering our strict selection criteria, our measures put a lower limit to the total amount of optically thick dust absorption in the universe. While we believe that most of the red-end flux excess in our galaxies can be attributed to optically thick dust absorption, quiescent/dormant galaxies or galaxies with emission lines whose equivalent width, by either AGN or star formation, is not properly captured in our SED modeling could also contribute \citep[e.g.,][]{2023ApJ...946L..16P}. We note, however, that in the targeted redshift range no bright lines are present in the bandpass we use for our search, namely the F444W band. Further observations with spectroscopy data will help us constrain the fraction of each contribution.

Our results strongly suggest that sizeable optically-thick dust absorption is very common if not ubiquitous in high-z massive galaxies with a wide variety of SFR, inclination angles, and mass concentration, making the partially dark galaxies indistinguishable from the fully UV/optical bright ones (non-candidates). Our sensitivity-driven lower limit to the stellar mass obscured by dark dust is 12\% of the total stellar mass. Furthermore, these results also show that optically thick obscuration in normal galaxies is a `local' feature, as opposed to an entire galaxy being made dark, and it could hardly affect the integrated properties of a galaxy unless dedicated searches such as the one presented here are conducted. In particular, the independence on the inclination angle suggests that dark substructures are randomly distributed inside galaxies so that projection effects could not change our estimates of the obscuration fraction statistically. We do find tentative evidence that galaxies with steeper UV spectral slopes are more likely to contain optically thick sub-regions. The specific spectral shape of the regions where the dark sub-regions are embedded varies, depending on the dilution by optically-think regions (and the details of the spatial decomposition procedure). But in our analysis, we have shown that they can still be recognizable and, on average, different when compared with fully UV/optical bright galaxies, i.e. with only optically thin obscuration.

Through spatial decomposition based on the F444W residual image (obtained by subtracting the model image photometrically scaled from the F277W observation), we are able to locate the optically dark substructures and study their properties. This turns out to be an efficient way to catch the differences in physical properties between UV/optical bright and dark sub-regions. We find that dark substructures tend to be more mass-concentrated and more asymmetric compared with the entire galaxy. There is a large variety in the morphology of the dark substructures, however. While a large fraction of them are circum-nuclear and relatively spherically symmetric, others are offset from the center and appear asymmetric. This variety of morphological properties helps explain why the amount of optically thick dust absorption is insensitive to the galaxies' inclination angles.

At z$>$3, the SFRD contributed by massive ($\log(M_*/M_{\odot})>$10.3) optically faint galaxies is at least two orders of magnitude higher than the contribution of Lyman-break galaxies with equivalent masses. By considering both optically faint galaxies and Lyman-break galaxies, the cosmic SFRD at z$=$4-5 is $\sim$43\% higher than that derived from UV-selected galaxies after dust correction\citep{2023A&A...672A..18X}. 

We find that one-third of normal massive galaxies at z$\sim$3 show evidence of optically thick dust obscuration, and around 10-20\% of the stellar mass or SFR is obscured in these galaxies. The optically-thick fraction is insensitive to either stellar mass or SFR in the selected mass range. Given our strict selection criteria, the total amount of optically thick dust absorption should be higher and we again stress that these numbers should be considered as lower limits. Also, probing galaxies at other redshifts could likely, we believe, result in different, i.e. larger, optically-thick fractions. Although the optically-thick fraction is relatively low in individual galaxies, the considerable fraction of candidate galaxies ($\sim$1/3 of the parent sample) makes its impact sizable. We conclude that the optically faint/dark substructures in normal galaxies significantly add to the cosmic SFRD and stellar mass density.

Another possibility that we have considered is whether the existence of optically faint/dark is a common transient evolutionary phase in the evolution of massive galaxies. \citet{2023ApJ...946....1C} claim that a significant fraction of star clusters in the mass range $0.3-2.5\times10^4 M_{\odot}$ remain within the dust cloud for at least 6 Myr before emerging from the clouds. If this is a common feature, it would indicate that optically dark sub-regions can last for a short but detectable timescale before clearing up the optically thick dust and emerging from the host galaxy. In this case, the morphology of the optically dark substructures, as well as the optically-thick fraction, would change as galaxies evolve. 
The correlation between the optically-thick fraction and recent burstiness (R1 and R2) reflects the timescales of dust formation and diffusion throughout the ISM in massive galaxies at z$\sim$3, where the physical conditions are significantly different from the local universe. At these redshifts, galaxies experience enhanced star formation rates ($\sim$one order of magnitude higher compared to local galaxies with similar stellar mass), larger gas fractions ($\approx$50\% and up to 80\%, see \citet{2020ARA&A..58..157T}), lower metal enrichment, higher ionization potential, and more compact morphologies, all of which are expected to affect the processes of formation of dust and its survival. There are likely marked differences in the dust properties between local galaxies and the ones we are considering, reflecting different types of chemical enrichment of the ISM. At $z\sim3$, the universe is $\approx2.1$ Gyr old, and dust production is dominated by the $\alpha$-elements generated by Type II SNe which enriched the ISM on very short timescales ($\sim$1 Myr, \citet{2003MNRAS.343..427M}). Later on, when Type Ia SNe explode, the ISM also gets enriched in Fe. The time delay between Type II and Type Ia SNe remains uncertain and can be of the order of the age of the universe at $z>2$ \citep{2016A&A...594A..54P, 2018MNRAS.479.3563F}. Meanwhile, dust mixing and homogenization within the ISM extend to $\sim$100 Myrs, varying with environmental factors like shock velocities and ISM density \citep{2011A&A...530A..44J}. While newly formed dust grains may quickly contribute to local obscuration, achieving sizable optically thick patches to be detected with current observations may require additional time. Given the complexity and uncertainty in modeling these dynamic processes at high redshifts, our analysis prioritizes empirical scaling relations rather than precise timescale estimates, highlighting correlations between dust properties and recent star formation bursts within the integrated galaxies. Our results suggest that optically thick dust absorption associated with local star-forming activities, in particular bursts, is quite common in the early evolution phase of massive galaxies. The emergence and disappearance of those dark substructures, as well as their correlation with the SFH, clearly need further investigation.

\section{Conclusion}
\label{sec:con}
We use JWST imaging data from the CEERS and PRIMER surveys to search for candidate localized optically-thick dust absorption in massive galaxies at z$\sim$3. Our main conclusions are summarized as follows:

\begin{itemize}

\item Optically thick dust absorption appears to be common, if not ubiquitous, in normal massive galaxies at 2.5$<$z$<$3.5 with $\log(M_*/M_{\odot})>$10.3. One-third of the galaxies in our sample have optically dark sub-regions, and these galaxies span a variety of SFR, inclination angles, and mass concentrations. Galaxies with steeply rising UV spectra are more likely to have optically dark substructures. 

\item Dark substructures can be recognized as dusty, mass-concentrated star-forming regions, irregularly distributed inside galaxies. They have a large variety of morphologies, with a big portion ($\sim$50\%) appearing compact and surrounding, but not necessarily concentric with, the galactic center.

\item Around one-third of the massive galaxies at z$\sim$3 show evidence of optically thick dust obscuration, which segregates on average $\sim$10-20\% of the stellar mass and SFR, i.e. they are invisible to UV/Optical observations. The optically-thick fraction is insensitive to stellar mass or SFRD in our sample. Thus, the total amount of optically thick dust absorption in our universe is larger than currently reported, making measures based on optically bright galaxies even stricter lower limits. 

\item The results presented here also suggest that optically thick dust absorption associated with localized star-formation activities inside galaxies could be a common evolutionary phase of massive galaxies. While our analysis suggests time scales between $\approx 30$ and $\approx 180$ Myr, further investigation is needed to understand the emerging and disappearing of the optically thick substructures, as well as their connection with the local, i.e. spatially resolved, star formation histories.

\end{itemize}

\bibliography{draft}{}
\bibliographystyle{aasjournal}

\appendix
\section{Criteria for candidate selection}
\label{a:cut}

\begin{figure*}
\plottwo{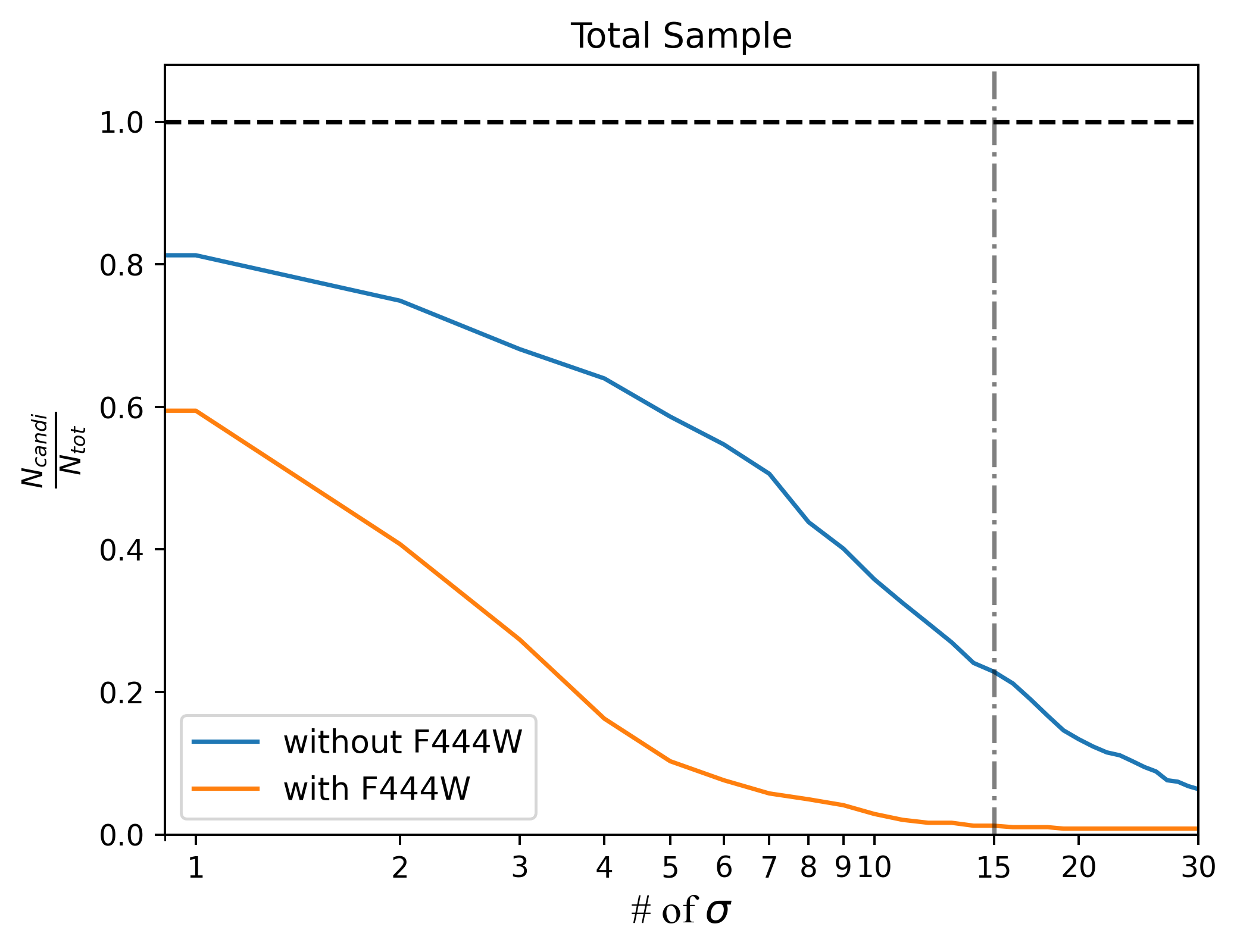}{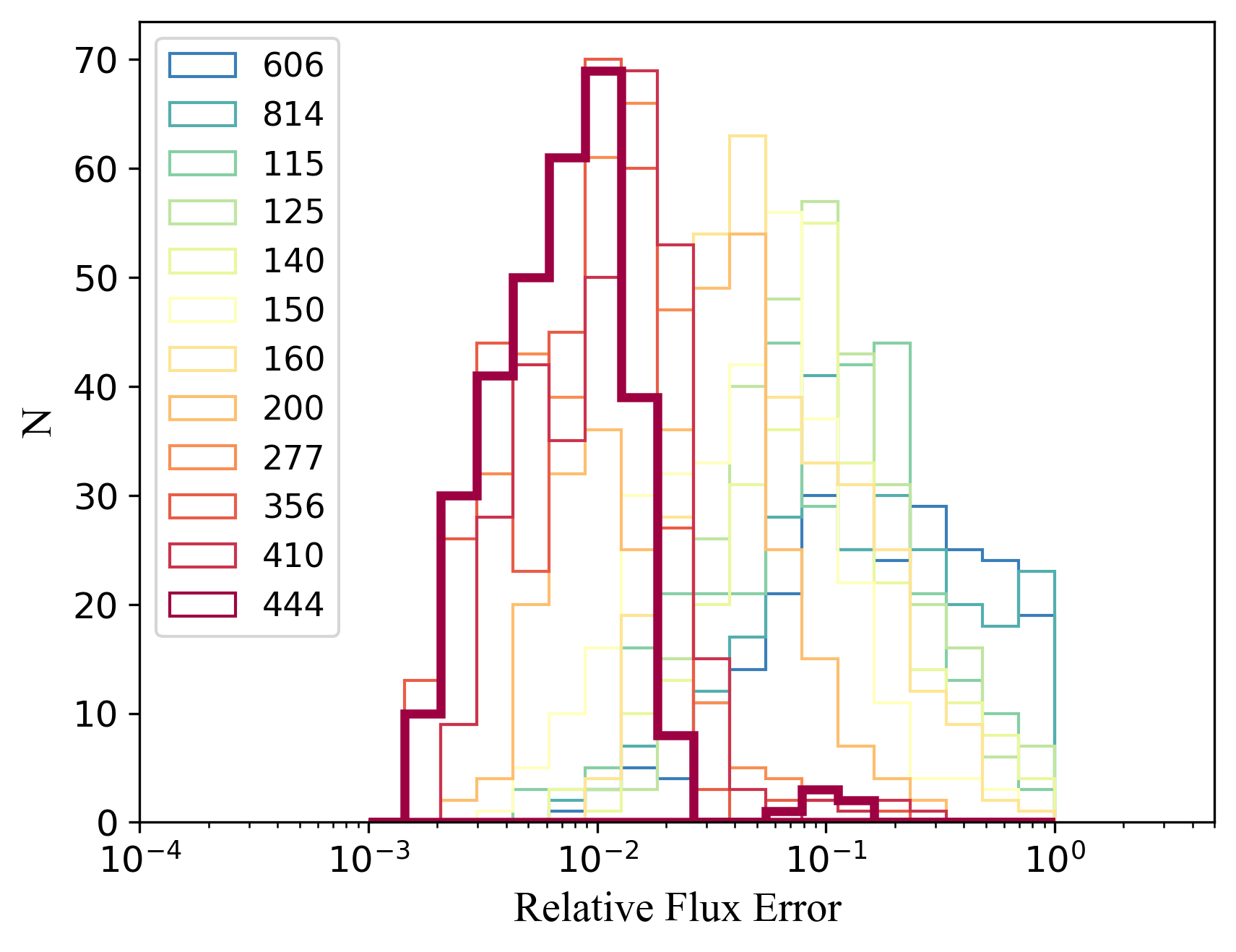}
\caption{Left: the number fraction of candidates as a function of the observed F444W band flux excess. The blue line refers to SED fittings without the F444W band photometry, while the orange line refers to fittings with the F444W band photometry. The final selection of $\sigma$ threshold cut is plotted as the gray vertical line. Right: the distribution of the relative photometric uncertainties for each filter. The F444W band is marked with a larger line width. \label{fig:select}}
\end{figure*}

As described in Section \ref{sec:candi}, we quantify the observed flux excess in the F444W band relative to the predicted one from SED modeling that does not include the passband. The excess is quantified by the number of $\sigma$ over the total uncertainty of the prediction (photometric uncertainty and posterior of the SED modeling procedure). The selection aims at picking up galaxies that have a robust excess, larger than a $\sigma$ threshold intentionally chosen to be large. We tracked the ratio of the number counts ratio of candidates to the total sample as a function of different $\sigma$ threshold cuts. As shown in the left panel of Figure \ref{fig:select}, when conducting SED fitting without the F444W band photometry (blue), the candidate fraction drops from $\sim80\%$ at 1$\sigma$ to $\sim10\%$ at 30$\sigma$. However, when conducting SED fitting with the F444W band (orange), the fraction is significantly lower in general and drops even more drastically. In particular, the orange curve drops to $\sim 0$ at around 15$\sigma$ while the blue curve stays above $20\%$. This is to say, no galaxy meets the criterion of 15$\sigma$ when including the F444W band in fitting, but we still get a significant number of candidates by excluding the F444W band in fitting. The observed flux excess in these galaxies is unlikely due to uncertainties in the SED modeling, and we regard them as solid candidates for the presence of optically thick dust absorption. Therefore, we adopt 15$\sigma$ as the threshold cut to pick up candidates from the parent galaxy sample.

The difference in the SED fitting results with or without the F444W band also depends on the weighting of the F444W data point (i.e., its photometric uncertainty relative to the other bands). As shown in the right panel of Figure \ref{fig:select}, the photometric uncertainties in F444W are systematically smaller than other bands with a narrow distribution, indicating that including F444W or not has a relatively significant impact on the fitting results.

\section{Validation of the spatial decomposition}
\label{a:decomp}

\begin{figure*}
\plotone{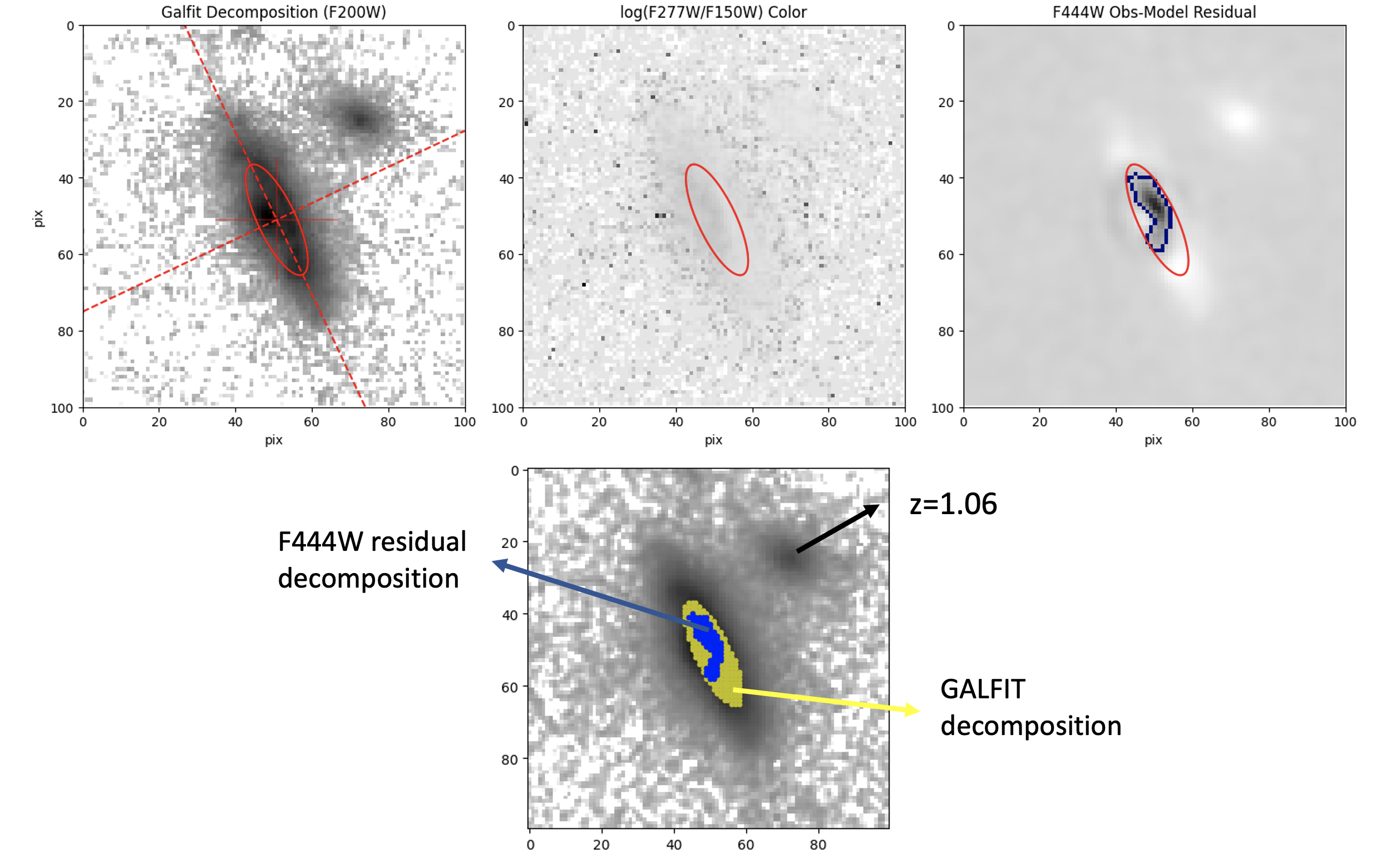}
\caption{The identification of dark substructures based on different decomposition methods for CEERS9-6044. Line 1: The red ellipse shows the bulge-disk decomposition given by \texttt{Galfit}. The F277W-F150W color map is shown in the middle panel, with the red ellipse overlaid to demonstrate the consistency. The blue irregular curve in the right panel marks the dark region identified by the F444W observation-model residual image. Line 2: different decomposition overlaid on the original F444W image. The nearby source is confirmed to be at lower redshift, being unrelated to the probed galaxy. \label{fig:decomp}}
\end{figure*}

In order to locate the location of the optically dark substructures within the candidate galaxies, we test three different spatial decomposition techniques.

We start with the traditional bulge-and-disk morphological decomposition. We adopt a bulge-disc decomposition scheme based on \texttt{Galfit} developed by the PHIBSS program \citep{2019A&A...622A.105F}. This method conducts two-component (bulge$+$disk) fits to the galaxies with a grid of initial guesses based on empirical studies, then the best-fit model is selected based on the lowest reduced chi-square. The decomposition result for an example galaxy (CEERS9-6044) is plotted in the first panel of Figure \ref{fig:decomp}.

Another way to approach the spatial decomposition is through the color maps of the galaxies, given that the regions that host the `dark regions' should appear redder than the rest of the galaxy. To mimic the rest-frame U-V colors, we use the F150W and F277W images to create a color map in logarithmic scale, as shown in the middle panel of Figure \ref{fig:decomp}. The red ellipse represents the bulge-disk decomposition given by \texttt{Galfit}. The region inside the ellipse appears darker, indicating a redder color. As a visual inspection shows, the morphological decomposition is highly consistent with the color decomposition.

Finally, we directly map the regions responsible for the F444W band flux excess by subtracting the model-predicted F444W image if the absorption inside the galaxies was only thin from the observed F444W image. The model-predicted image is derived by normalizing the observed F277W image according to the best-fit SED model. The observation$-$model residual image is shown in the right panel of Figure \ref{fig:decomp}. We use a 10$\sigma$ contour (dark blue) to outline the residual flux where in this case $\sigma$ is the local background noise. The bulge-disk decomposition is also overlaid as a red ellipse. It turns out that the irregular dark blue region is well within the ellipse, precisely identifying the reddest portion of the galaxy.

To summarize, we find high consistency among different spatial decomposition techniques. To maximize the efficiency in isolating the dark patches, we adopt the F444W band `Obs$-$Model' residual image approach to define the dark substructures within candidate galaxies.

\section{SFHs with different burstiness}
\label{a:sfh_burst}

\begin{figure*}
\plotone{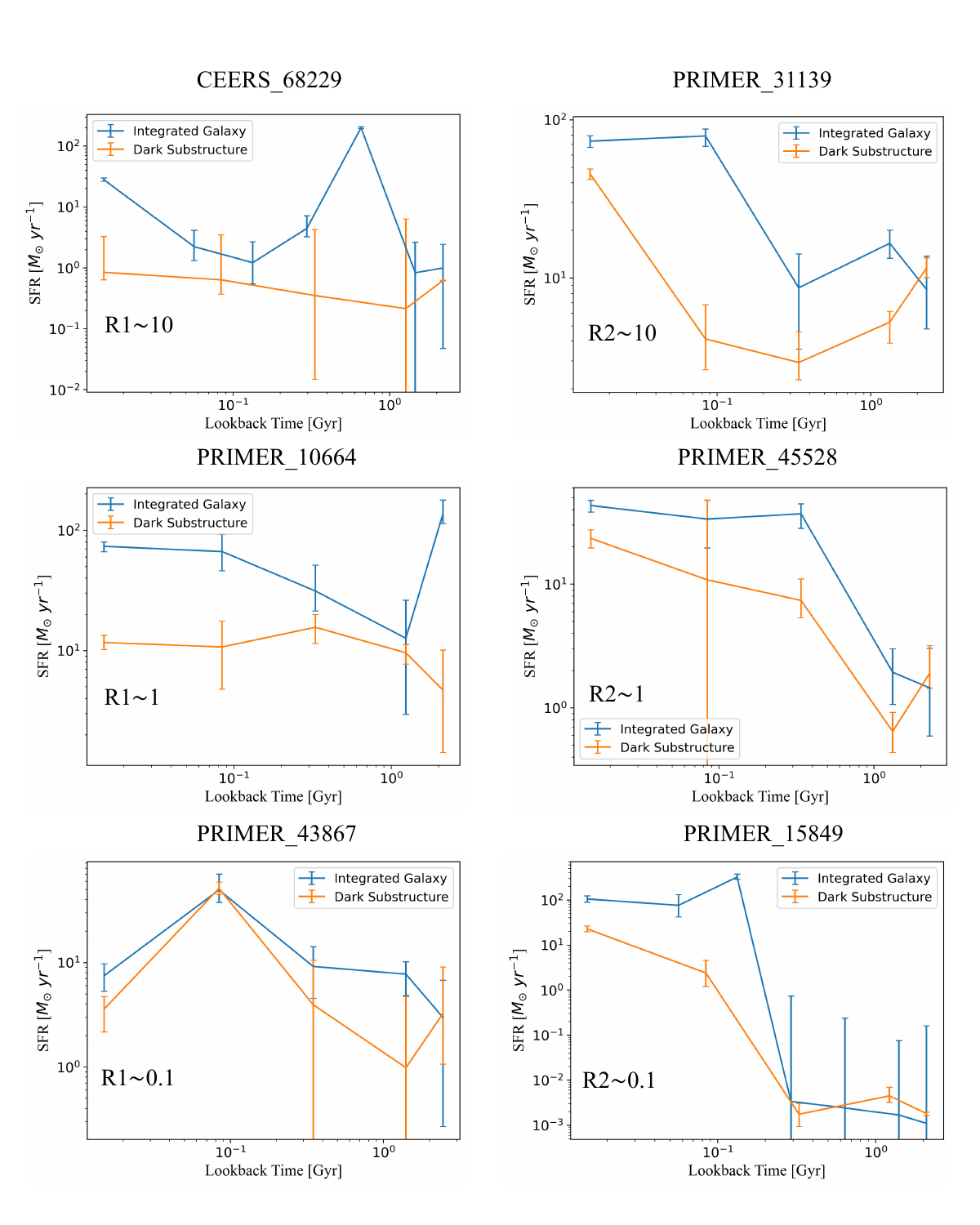}
\caption{Here we show a few examples of the SFHs given by \texttt{Prospector} fit with different burstiness measurements. The burstiness is defined by the SFR ratio in certain timescales, where $R1=\frac{SFR_{0-30~Myr}}{SFR_{30-80~Myr}}$, and $R2=\frac{SFR_{0-80~Myr}}{SFR_{80-180~Myr}}$. The SFH of the integrated galaxy is plotted in blue, and the SFH of the dark substructure is plotted in orange. \label{fig:sfh_example}}
\end{figure*}

\end{document}